\documentclass[useAMS,usenatbib]{mnras}
\usepackage{graphicx,amsmath,color,amssymb}
\usepackage[pdftitle={Statistical properties of DLAs from SDSS DR12},pdfauthor={Simeon Bird}]{hyperref}	
\hypersetup{colorlinks=true,linkcolor=blue,citecolor=blue,filecolor=blue,urlcolor=blue}
\usepackage[all]{hypcap}
\usepackage[T1]{fontenc}
\usepackage{ae,aecompl}

\newlength{\narrowFigurewidth}
\newlength{\Figurewidth}
\newlength{\wideFigurewidth}
\def\jcap{JCAP}        

\newcommand{\Lya}{Lyman-$\alpha$}

\newcommand{\NHunit}{cm$^{-2}$}

\newcommand{\NHI}{N_\mathrm{HI}}

\title[Statistical properties of DLAs from SDSS DR12]{Statistical properties of damped Lyman-alpha systems from Sloan Digital Sky Survey DR12}

\author[S.\ Bird et al.]{Simeon Bird,$^1$\thanks{E-mail: sbird4@jhu.edu} Roman Garnett,$^2$ Shirley Ho$^3$ $^4$ $^5$\\
  $^1$Department of Physics \& Astronomy, Johns Hopkins University, 3400 N.\ Charles Street, Baltimore, MD 21218, USA\\
  $^2$Department of Computer Science and Engineering, Washington University in St.\ Louis, One Brookings Drive, St.\ Louis, MO 63130, USA\\
  $^3$Department of Physics, Carnegie Mellon University, 5000 Forbes Avenue, Pittsburgh, PA 15213, USA\\
  $^4$Lawrence Berkeley National Laboratory, 1 Cyclotron Road,Berkeley, CA 94720, USA \\
  $^5$Department of Physics, University of California , Berkeley, CA 94720 \\
}

\begin{document}

\maketitle

\begin{abstract}
We present new estimates for the statistical properties of damped Lyman-$\alpha$ absorbers (DLAs).
We compute the column density distribution function at $z>2$, 
the line density, $\mathrm{d}N/\mathrm{d}X$, and the neutral hydrogen density, $\Omega_\mathrm{DLA}$.
Our estimates are derived from the DLA catalogue of \cite{Garnett:2016}, which uses the SDSS--III DR12 
quasar spectroscopic survey. This catalogue provides a probability 
that a given spectrum contains a DLA. It allows us to use even the noisiest data without biasing 
our results and thus substantially increases our sample size.
We measure a non-zero column density distribution function at $95\%$ confidence for all 
column densities $\NHI < 5\times 10^{22}$ \NHunit. We make the first measurements from SDSS of 
$\mathrm{d}N/\mathrm{d}X$ and $\Omega_\mathrm{DLA}$ at $z>4$.
We show that our results are insensitive to the signal-to-noise ratio of the spectra, but that 
there is a residual dependence on quasar redshift for $z<2.5$, which may be due to remaining 
systematics in our analysis.
\end{abstract}


\section{Introduction}
\label{sec:intro}

Damped \Lya~absorbers (DLAs) are absorption systems corresponding to
neutral hydrogen column densities $> 2\times 10^{20}$ \NHunit\  \citep{Wolfe:1986}.
Observed primarily in quasar spectra at $z>2$, they arise from gas dense enough
to self-shield from the effect of the ultra-violet background (UVB) \citep{Cen:2012}, but diffuse enough
to have a low star formation rate \citep{Fumagalli:2015}. DLAs dominate the neutral
hydrogen budget of the Universe after reionisation \citep{Gardner:1997, Noterdaeme:2012, Zafar:2013, Crighton:2015}. 
Simulations indicate that they are connected
with a wide variety of halo masses, peaking in the range $10^{10}$--$10^{11} M_\odot$
\citep{Haehnelt:1998, Prochaska:1997, Pontzen:2008}. The DLA halo mass can also be inferred by measuring the 
clustering of DLAs, through their cross-correlation with the \Lya~forest \citep{FontRibera:2012}. 
Present clustering data indicates a halo mass moderately larger than anticipated by simulations \cite[although see][]{Bird:2014a}.

Metal absorbers have been detected associated with DLAs \citep{Rafelski:2012}.
DLAs show a wide range of metallicity, decreasing at higher redshift, 
with a median value which is subsolar \citep{Jorgenson:2013, Rafelski:2014, MasRibas:2016}. The DLA 
metallicity is expected to be correlated with the host halo 
mass \citep{Pontzen:2008}.
DLAs show little dust reddening \citep{Khare:2012} and a small subset are 
amongst the most metal-poor objects known
\citep{Pettini:2008}. Furthermore, DLAs are usually hosted by low luminosity objects, whose low star 
formation rate means they are not currently observable in emission \citep{Fumagalli:2015}.

In this work we present new estimates for the abundance of DLAs, the average neutral hydrogen density,
and the column density distribution function at $z>2$.
These population statistics are important in the wider context of galaxy formation theory.
As neutral hydrogen is the raw material for star formation, measurements of neutral hydrogen can provide
an independent constraint on theoretical models tuned to match the galaxy stellar mass
function.

We compute statistics using the DLA catalogue presented in \cite{Garnett:2016}\footnote{Available at \url{http://tiny.cc/dla_catalog_gp_dr12q_lyb}}, an
analysis of Sloan Digital Sky Survey III Data Release 12 (SDSS--III DR12) \citep{Eisenstein:2011, Dawson:2013, Alam:2015}
using a set of statistical techniques based on machine learning and Gaussian processes. DR12 contains
the largest currently available quasar spectroscopic survey, but to achieve throughput it
sacrifices spectrograph resolution and has relatively low signal-to-noise ratio in typical measurements. These properties make
detecting DLAs challenging, despite the distinctive shape of the damping wing of the Voigt profile, which is 
a powerful discriminant from other absorbers.
Nevertheless, several methods have been used to successfully detect DLAs in SDSS spectra.
These have included a visual-inspection survey \citep{slosar_et_al_jcap_2011},
visually guided Voigt profile fitting \citep{Prochaska:2005, Prochaska:2009},
a template-matching approach \citep{Noterdaeme:2009, Noterdaeme:2012}
and an unpublished machine-learning approach using Fisher discriminant analysis \citep{carithers_unpublished_2012}.
These methods together have succeeded in detecting large catalogues of DLAs, but each relies on
a labour-intensive visual component that limits their ability to scale to larger surveys.
The methods described in \cite{Garnett:2016}, which use these prior catalogues as training data,
are fully automated, allowing them to be used even for forthcoming much larger surveys without guidance.

The DLA catalogue of \cite{Garnett:2016} is for the first time able to provide a probability
of the presence of a DLA in each quasar spectrum, allowing for ``soft'' detections in noisy data. Furthermore, the catalogue includes a full posterior distribution for the column density
and redshift of a putative DLA. This allows us to propagate uncertainties from the inference performed on each spectrum into
the global population. In particular, we are often uncertain as to whether a particular low signal-to-noise spectrum
contains a DLA. This will be reflected in the computed probability of that spectrum containing a DLA, allowing us to account for this uncertainty without biasing
our final results. Thus we are able to include the full sample of low signal-to-noise DLAs in our measurements,
substantially increasing the sample size.

Section \ref{sec:methods} details how we estimate the statistical properties of the DLAs from the catalogue.
Then in Section \ref{sec:results} we present our results. Section \ref{sec:cddf} discusses the column density
distribution function (CDDF), while Section \ref{sec:dndx} gives the line abundance of DLAs per unit
pathlength and the average universal density in neutral hydrogen. We present null checks for systematic
error in Section \ref{sec:systematic}; Section~\ref{sec:snr} shows independence of our result to
signal-to-noise ratio, while Section \ref{sec:bright} discusses a residual sensitivity
to quasar redshift. We conclude in Section \ref{sec:conclusions}.

\section{Methods}
\label{sec:methods}

DLAs are detected using the method presented in \cite{Garnett:2016}. This method
uses Bayesian model selection, with priors derived from SDSS DR9 \citep{Lee:2013}. 
It produces a probability that each DR12 spectrum contains a DLA as well as a full posterior probability 
distribution for the DLA redshift and column density in the case that it does contain a DLA. Here we 
present a pedagogical summary of the approach, deferring to \cite{Garnett:2016} for
further details. All our code is publically available at \url{https://github.com/rmgarnett/gp_dla_detection}, and 
the DLA catalogue can be found at \url{http://tiny.cc/dla_catalog_gp_dr12q_lyb}.

As DLAs are observed via absorption in quasar spectra, we need a model for the emission properties of a quasar in the absence of a DLA.
This spectrum includes the quasar continuum, observational noise, and associated absorption from matter between the quasar and the observer.
We model all these quantities using Gaussian processes \citep{gpml}.
A Gaussian process is a nonparametric distribution over functions, the function space analogue of the normal distribution.
To use Gaussian processes to make predictions for new data, it is necessary to impose a prior form for the mean emission and the covariance between different observational data points.
Most applications of Gaussian processes use a standard, off-the-shelf form for the covariance, such as the exponential squared kernel.
However, given the complex shape of quasar continua, no such standard form is appropriate. Instead, we learn a prior distribution for the shape of quasar emission without a DLA
using a training sample of previously analysed quasar spectra, for which we already have DLA information.
Specifically, we trained our model on the concordance DLA catalogue from SDSS DR9 \citep{Lee:2013}. This concordance DLA catalogue is the ``majority vote'' of the DLA catalogue
published as \cite{Noterdaeme:2009}, an unpublished DLA catalogue by \cite{carithers_unpublished_2012}, and DLAs found by visual inspection \citep{slosar_et_al_jcap_2011}.
Overall, it comprises a large sample, containing $\sim 50,000$ spectra. This training set enables us to estimate the prior probability and covariance of a given DLA-free 
quasar spectrum, which we model with a Gaussian process with mean and covariance functions we fit to the data.

We model both observational noise and absorption from matter along the path between the observer and the quasar by adding
Gaussian noise contributions with appropriate variances to the emission model. The variance for the observational noise model is a function of the (observer-frame) wavelength of the spectrum,
and given by the SDSS pipeline noise estimates. The variance for the forest absorption, corresponding to the mean flux from absorbing material, is learnt
in a similar way to the emission model.

We may then transform the derived model for DLA-free quasar spectra into a model for
the shape of an observed quasar spectrum containing a DLA.
Gaussian processes offer a straightforward mechanism to incorporate simulated absorption profiles of a putative DLA. Each DLA is parametrised
by its redshift and column density; given these parameters, we may evaluate the likelihood of the data under the alternative DLA model.  
This likelihood will be high when the simulated
absorption profile explains the observed data better than the simpler DLA-free model. To compute the evidence of this model given the data, we must marginalise the a priori unknown 
DLA parameters under a chosen prior distribution. The DLA redshift is a priori assumed to be distributed
uniformly between a point $3000$ km/s blue-wards of the quasar emission redshift (to avoid a possible
quasar proximity zone) and the wavelength of Lyman-$\beta$ emission in
the quasar rest frame (to avoid absorption from the Lyman-$\beta$ forest)\footnote{Note that this differs
slightly from the catalog of \cite{Garnett:2016}, who used the wavelength of the Lyman limit as a lower bound.}.
The column density, $\NHI$, has a prior distribution given by a mixture between the
column density function from the training set and, to allow for measurement uncertainty, a uniform prior on $\log\NHI$.

We now have two models for the observed spectrum of the quasar, one modelling the emission in the absence of a DLA, the other in the presence of a DLA. We would like to compare how well they explain a given spectrum,
and thus find the posterior probability of a DLA with a particular column density or redshift along the line of sight. To do this we evaluate the Bayesian evidence
for each model, assuming that the prior probability of a DLA is uniform in quasar redshift path length, and
equal to the overall fraction of DLAs per unit redshift path in the training set. This integral is non-analytic, so we estimate it via numerical quadrature with 10\,000 realizations of DLA parameters covering the space exhaustively.
The posterior distribution over DLA parameters may be derived via normalization.

Our ability to obtain the full posterior probability distribution for the properties of a putative DLA is a strong advantage of our technique when estimating derived properties of the DLA population.
We are able to use this posterior distribution to draw inferences about the properties of the overall DLA population even from
quite noisy data, which may leave considerable uncertainty in the properties of any particular DLA or quasar. We will propagate this uncertainty through the analysis, ultimately allowing us to derive stronger statistical power than was available previously.

Refinements in future work might involve modifying our analysis pipeline to include data from metal lines associated with the DLA and thus better constrain the redshift and column densities of individual absorbers.

\subsection{Estimating DLAs in a Column Density Bin}

In this section, we explain how we turn our measured posterior distribution of DLAs, determined as explained
above and in \cite{Garnett:2016},
into a measurement of the average binned column density distribution function (CDDF),
the abundance of DLAs with redshift ($\mathrm{d}N/\mathrm{d}X$) and average matter density in DLAs ($\Omega_\mathrm{DLA}$).
To do this we will first compute the expected number of DLAs in a given column density and redshift bin. This involves computation
of summed products of probabilities from all spectra within the sample. We will show that this computation can be done directly with
a few simple approximations.

The DLA pipeline provides, for each spectrum, the probability that it contains a DLA, $p_\mathrm{DLA}^i$, and
the likelihood of each sample in the $(\NHI, z)$ space.
The probability of a DLA in a given redshift and column density range in a particular spectrum is the sum of
the likelihoods for all samples in the range, multiplied by the probability of the spectrum containing a DLA.

The total number of DLAs in the entire spectral sample in a given bin, $N$, is the sum of $n$ binomial processes (the Poisson--binomial distribution),
and so the probability that there are $N$ DLAs is:
\begin{equation}
    P(N) = \sum_{n=N} \prod_{i \in \mathrm{DLA}} p_\mathrm{DLA}^i \prod_{i \not \in \mathrm{DLA}} (1-p_\mathrm{DLA}^i)\,.
    \label{eq:dlalikelihood}
\end{equation}
Here $p_\mathrm{DLA}^i$ is the probability that the $i$th spectrum contains a DLA. The sum is over all possible choices of $N$ DLAs from $n$ possible spectra. For each choice,
there are $N$ spectra that contain a DLA and $n-N$ which do not. The first product is over the $N$ spectra which
do contain a DLA in this draw of $N$ DLAs from $n$ spectra. 
The second product is over all other spectra, the $n-N$ which do not contain a DLA in this draw.

For the large number of spectra we consider here, it is too numerically intensive to evaluate this function directly.
\cite{Lecam:1960} showed that the Poisson--binomial distribution can be approximated as a Poisson distribution for small $p$.
To take advantage of this, we split the spectra into two sets, based on whether the probability of a DLA
in this bin is greater or less than a transition probability, $p_\mathrm{switch}$.
If $p < p_\mathrm{switch}$, we approximate the Poisson--binomial as a Poisson distribution
with mean $\sum p_\mathrm{DLA}^i$. For spectra with $p > p_\mathrm{switch}$,
Eq.\ \ref{eq:dlalikelihood} is evaluated directly using a discrete Fourier transform \citep{Fernandez:2010}.
The two probability distributions are then combined and the confidence intervals computed.
Following \cite{Lecam:1960}, we use $p_\mathrm{switch} = 0.25$, but we have verified that
our results are unchanged with $p_\mathrm{switch} = 0.5$. We also
neglect (that is, assume they never contain a DLA) spectra where $p_\mathrm{DLA}^i < 0.05$, as well as
individual samples with $p_\mathrm{DLA}^i < 10^{-4}$, as
an optimisation. We have verified that lower probability spectra and samples make a negligible
difference to our results.

This procedure allows us to compute the posterior probability
distribution of the total number of DLAs in the sample. The confidence intervals thus reflect the
uncertainty from DLA detections and redshift and column density estimation.
There is also a sampling error, arising from the finite sampling of space by quasar
sightlines. However, due to the large sample provided by SDSS--III, this error is very small.
We have verified this by resampling our catalogue multiple times with replacement (bootstrap errors).
The central value and spread of the results was extremely similar to the full confidence limits, verifying
that sample variance is indeed subdominant.

We define the column density distribution, $f(N)$, such
that $f(N)$ is the number of absorbers per unit column density per unit absorption distance with column density
in the interval $[N, N + {\rm d}N]$. Thus
\begin{align}
 f(N) &= \frac{F(N)}{\Delta N \Delta X(z)}\,,
\end{align}
where $F(N)$ is the number of absorbers along simulated sightlines in a given column density bin. $\Delta X(z)$ is the absorption distance per sightline,
defined to account for evolution in line number density with the Hubble flow:
\begin{equation}
 X(z) = \int_0^z (1+z')^2  \frac{H_0}{H(z')} \,{\rm d}z'\,.
 \label{eq:absdist}
\end{equation}
Here the Hubble function is $H^2(z)/H^2_0 = \Omega_\mathrm{M} (1+z)^3 + \Omega_\Lambda$.

We also show the incident rate of DLA systems, $\frac{\mathrm{d}N}{\mathrm{d}X}$, and
the total HI density in DLAs, $\Omega_\mathrm{DLA}$.
$\frac{\mathrm{d}N}{\mathrm{d}X}$ is the integral of the CDDF:
\begin{equation}
 \frac{\mathrm{d}N}{\mathrm{d}X} = \int_{10^{20.3}}^\infty f(N, X) \,\mathrm{d} N\,,
 \label{eq:dndx}
\end{equation}
while $\Omega_\mathrm{DLA}$ is the integral of its first moment:
\begin{equation}
 \Omega_\mathrm{DLA} = \frac{m_\mathrm{P} H_0}{c \rho_c}\int_{10^{20.3}}^\infty N f(N, X) \,\mathrm{d} N\,.
 \label{eq:omdla}
\end{equation}
Here $\rho_c$ is the critical density at $z=0$ and $m_\mathrm{P}$ is the proton mass. Note that this definition
differs by a factor of $X_H = 0.76$, the primordial hydrogen mass fraction, from the quantity quoted by \cite{Noterdaeme:2012}.
The difference comes from the fact that they divide the total mass of neutral hydrogen by the hydrogen mass fraction
in order to obtain the total gas mass in DLAs. However, this assumes the neutral fraction of hydrogen is unity,
while in fact some of the hydrogen in DLAs may be molecular, implying that total gas mass in DLAs
is greater than $\Omega_\mathrm{DLA} / 0.76$ by an uncertain amount. We therefore quote the total HI mass, and have
adjusted the results from \cite{Noterdaeme:2012} to match our convention.
Because the CDDF falls off sharply above the DLA threshold, $\frac{\mathrm{d}N}{\mathrm{d}X}$ is dominated by
systems with $\NHI \sim 5\times 10^{20}$ \NHunit, while $\Omega_\mathrm{DLA}$ is strongly affected by higher column
densities with $\NHI \sim 10^{21}$ \NHunit.

Note that the \cite{Noterdaeme:2012} measurements of $\frac{\mathrm{d}N}{\mathrm{d}X}$ do not include error bars,
as the error was dominated by the uncertainty in detections of weak DLAs (P. Noterdaeme,
private communication). An advantage of our method is that we can quantify this uncertainty,
and thus we do include error bars on $\frac{\mathrm{d}N}{\mathrm{d}X}$.

We omit from our catalogue the regions of spectra potentially affected by Lyman-$\beta$ absorption, that is,
those with a rest wavelength of $\lambda < 1\,026.72 $~\AA. As our simple model for the \Lya~forest absorption
did not include Lyman-$\beta$, the Gaussian process code misidentifies some of this
absorption (especially in the most noisy spectra) as potential DLAs. The redshift distribution of SDSS quasars peaks at
relatively low redshift, meaning that the total fraction of path length affected is relatively small. Thus we simply
regenerate the catalog while excising the affected part of the spectrum. An alternative method would be to explicitly model absorption from Lyman-$\beta$, something we may attempt in future work.

We assume that each spectrum contains at maximum one DLA. By neglecting spectra containing two strong DLAs, this
may marginally decrease the $\mathrm{d}N/\mathrm{d}X$ over that observed.  The most detectable DLA in a spectrum is likely to occur
at low redshift, so the objects missed will be more common at high redshift. However, DLAs are relatively
rare (and the spectrum we examine limited to that between $1\,216$ and $1\,027$ \AA) so this is unlikely to be a significant effect. Note this
only matters for spectra containing two objects that are overwhelmingly likely to be DLAs, ie, where the expected
number of DLAs in the spectrum is greater than one. A spectrum containing two objects with $p_{\mathrm{DLA}} \sim 0.1$, for example,
will be counted correctly as $p_{\mathrm{DLA}} \sim 0.2$.

When computing $\Omega_\mathrm{DLA}$, we must compute not only the number of DLAs
in a redshift bin, but also their total column density. For computational reasons, we
estimate this by constructing the column density function in each redshift bin of interest and then using Eq.~\ref{eq:omdla}
to obtain $\Omega_\mathrm{DLA}$ by integration. For each redshift bin we compute the full posterior PDF for $30$ column density bins, using the method described above. These (discrete) PDFs are then multiplied together
to estimate the posterior distribution for the total matter density at each redshift.

\section{Results}
\label{sec:results}

\subsection{Column Density Distribution Function}
\label{sec:cddf}

\begin{figure}
\includegraphics[width=0.45\textwidth]{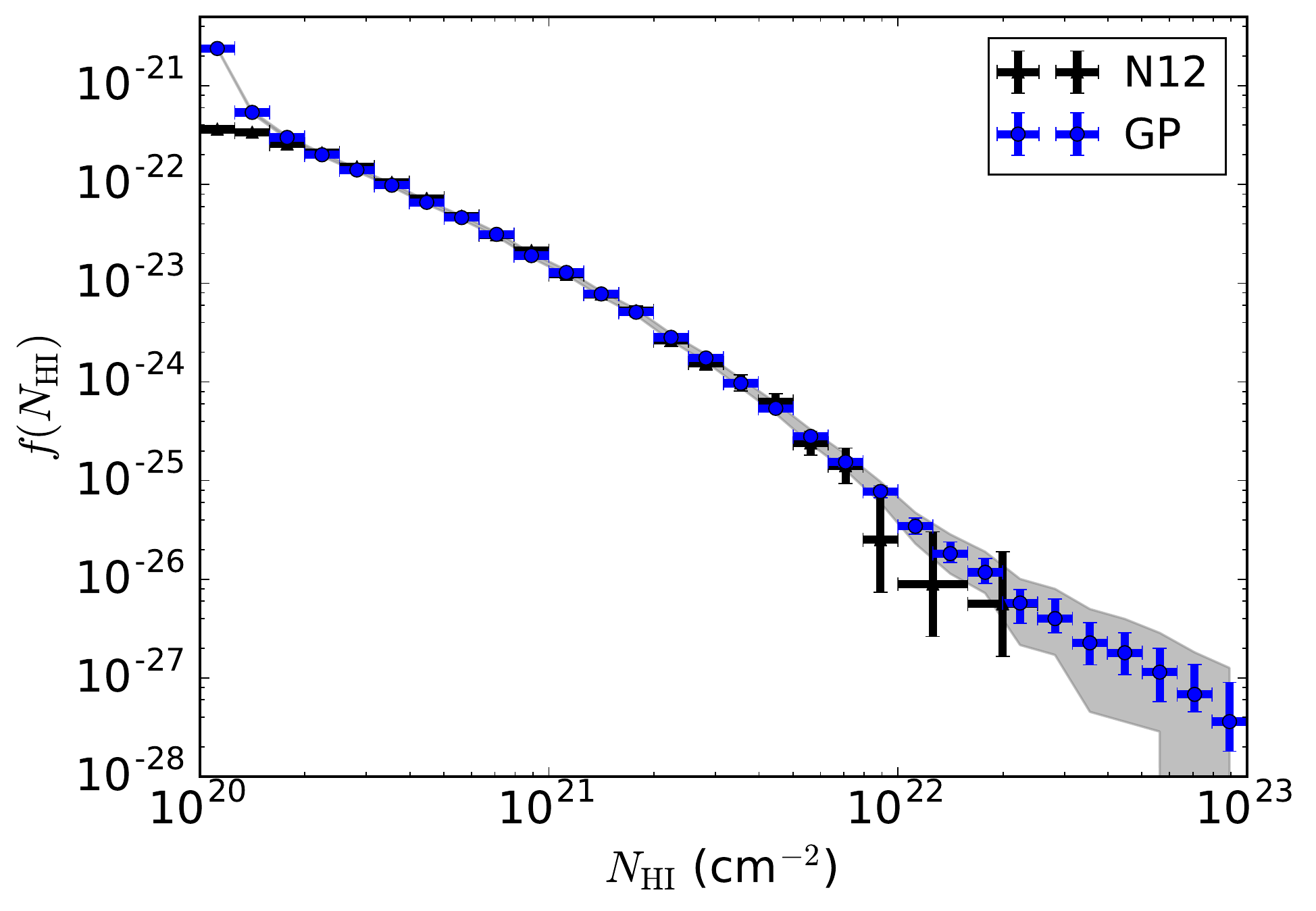}
\caption{
The CDDF, integrated over all $z<5$ spectral path, derived from our DR12 DLA catalogue (blue, labelled `GP')
and compared to the CDDF derived from the DLA catalogue of \protect\cite{Noterdaeme:2012} (N12; black). 
$68\%$ confidence limits are shown as error bars, while $95\%$ confidence limits are shown as a grey filled band. Note that at low column densities our results completely overlap those of N12, and the large sample means that errors are extremely small.
}
\label{fig:cddf}
\end{figure}

\begin{figure}
\includegraphics[width=0.45\textwidth]{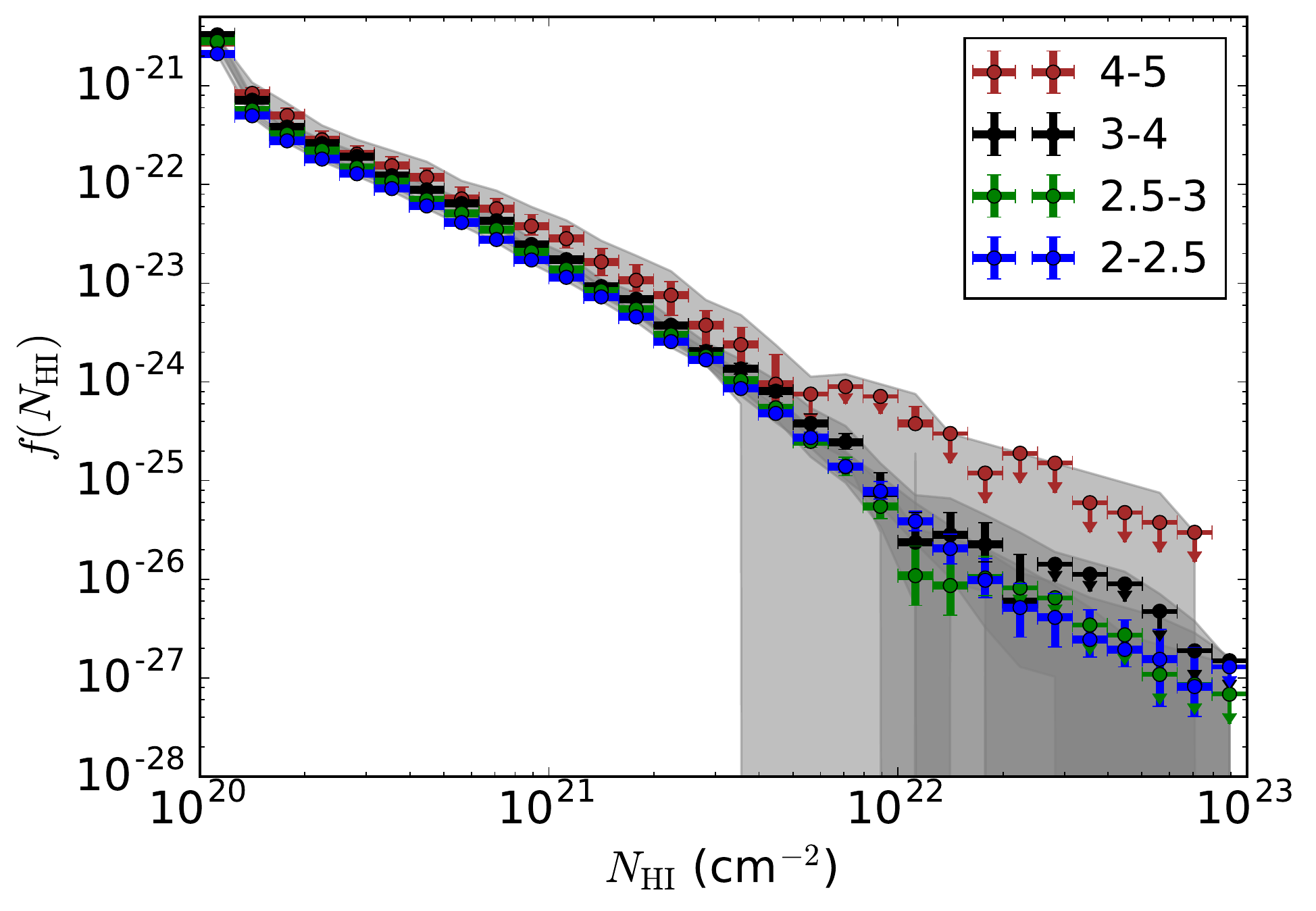}
\caption{
The CDDF derived from DLAs in a variety of redshift ranges. Labels show the redshift range of the absorbers used.
$68\%$ confidence limits are shown as error bars, while $95\%$ confidence limits are shown as a grey filled band. In
column density and redshift ranges where there was no detection at $68\%$ confidence, a down-pointing arrow is shown
indicating the $68\%$ confidence upper limit.
}
\label{fig:cddf_zz}
\end{figure}

Figure \ref{fig:cddf} shows the CDDF from our analysis of DR12,
compared to the DR9 DLA catalogue from \cite{Noterdaeme:2012}. We combine
all spectral path where an absorber would be at $z<5$ to obtain a path-weighted average over redshift. In practice for SDSS DR12 the total path
is dominated by $z<2.5$. Error bars denote 68\% confidence limits, while the grey band shows the region enclosed by the 95\% confidence limits.
Table \ref{tab:DR12/cddf_all.txt} contains our CDDF in tabulated form.

The lowest two column density bins are somewhat high due to an artefact of
our analysis method. Our column density prior for each spectrum is cut off at $\NHI < 10^{20}$,
but the lower limit is purely a matter of the conventional definition of a DLA. Our detection method is capable of
distinguishing somewhat weaker absorbers from noise, and correctly assigns them a maximum
posterior column density corresponding to the lowest value allowed by the prior. Note this will
not affect our results for $\mathrm{d}N/\mathrm{d}X$ or $\Omega_\mathrm{DLA}$, because by definition these
only include DLAs, i.e., samples with $\NHI > 10^{20.3}$. We may perform further analysis
of these weak absorbers in future work, effectively extending our prior to lower column densities.

Our results agree well with those of \cite{Noterdaeme:2012}. Our error bars are substantially smaller,
emphasising the increased statistical size of our sample. This is due not just to the larger size of DR12,
but also to the ability of our method to quantify the uncertainty in a DLA measurement. This allows us to robustly include
noisier spectra into our estimate, greatly increasing the size of the available sample.
Interestingly the data shows no indication of a molecular hydrogen cutoff at high column density \citep{Altay:2011}.
We detect the presence of DLAs at 95\% confidence for $\NHI < 6 \times 10^{22}$~\NHunit, and at 68\% confidence for $\NHI < 10^{23}$~\NHunit.

Figure \ref{fig:cddf_zz} shows the CDDF in a variety of redshift bins. Symbols indicated by downward pointing arrows
show the $68\%$ upper confidence limit when the data is consistent with zero at $68\%$ confidence.
We detect minimal evolution in the shape of the CDDF with redshift for $z< 4$. The CDDF amplitude gently increases with redshift, as we shall discuss in
more detail by looking at $\mathrm{d}N/\mathrm{d}X$ and $\Omega_\mathrm{DLA}$ in Section \ref{sec:dndx}. Interestingly, there is some evidence of the slope of the CDDF
becoming shallower at $z>4$, although we do not yet reliably detect DLAs with column density $\NHI > 3 \times 10^{21}$~\NHunit.
It will be interesting to see if larger data sets confirm a shallower slope, or point to a cutoff.

\subsection{Redshift Evolution of DLAs}
\label{sec:dndx}

\begin{figure}
\includegraphics[width=0.45\textwidth]{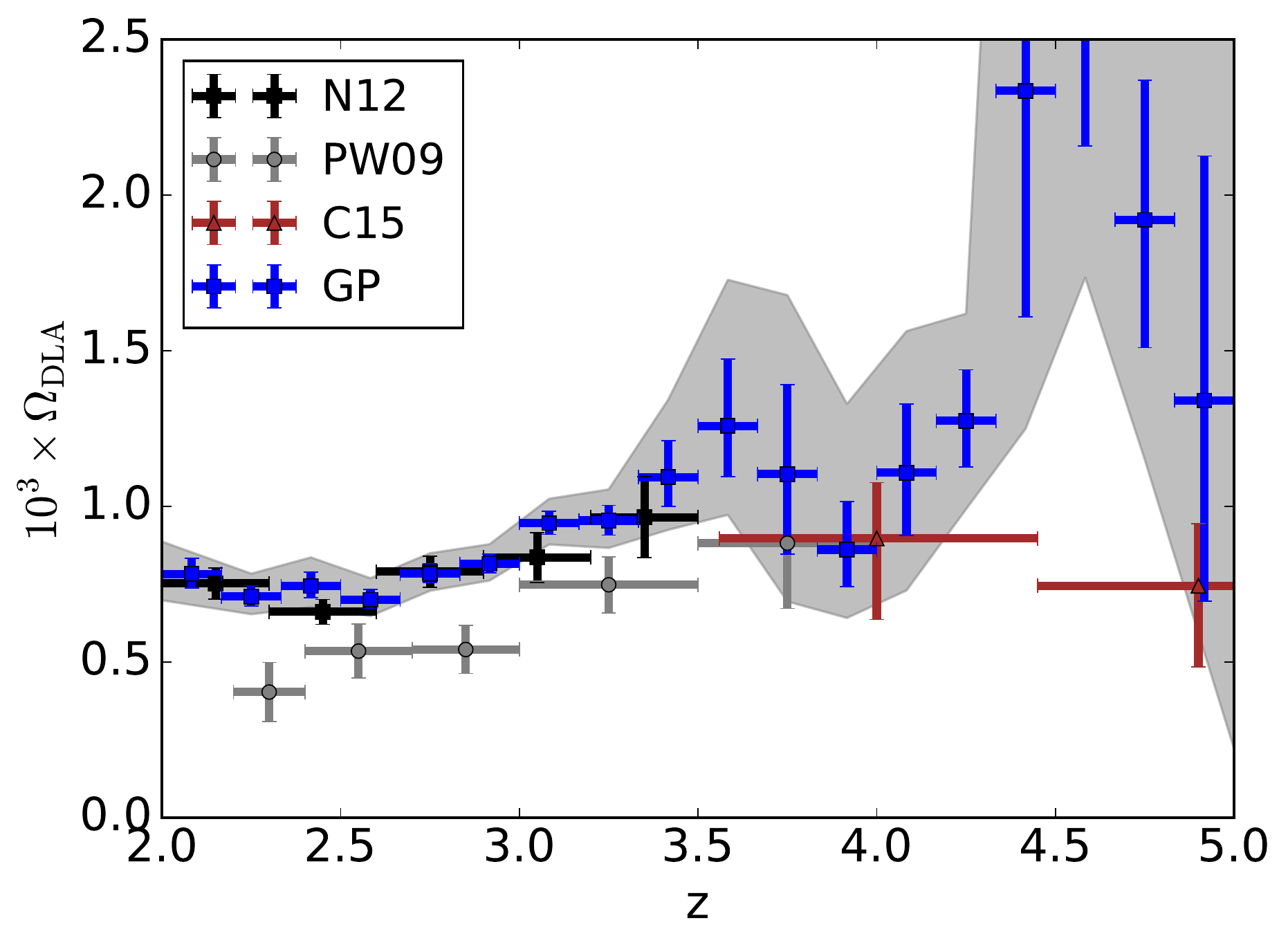}
\caption{
The HI density in DLAs, $\Omega_\mathrm{DLA}$, from our DR12 DLA catalogue as a function of redshift (blue, labelled `GP'),
compared to the results of \protect\cite{Noterdaeme:2012} (N12; black), \protect\cite{Prochaska:2009} (PW09; grey) and \protect\cite{Crighton:2015} (C15; red). See text for an explanation of the mild tension between measurements.
}
\label{fig:omega_dla}
\end{figure}

\begin{figure}
\includegraphics[width=0.45\textwidth]{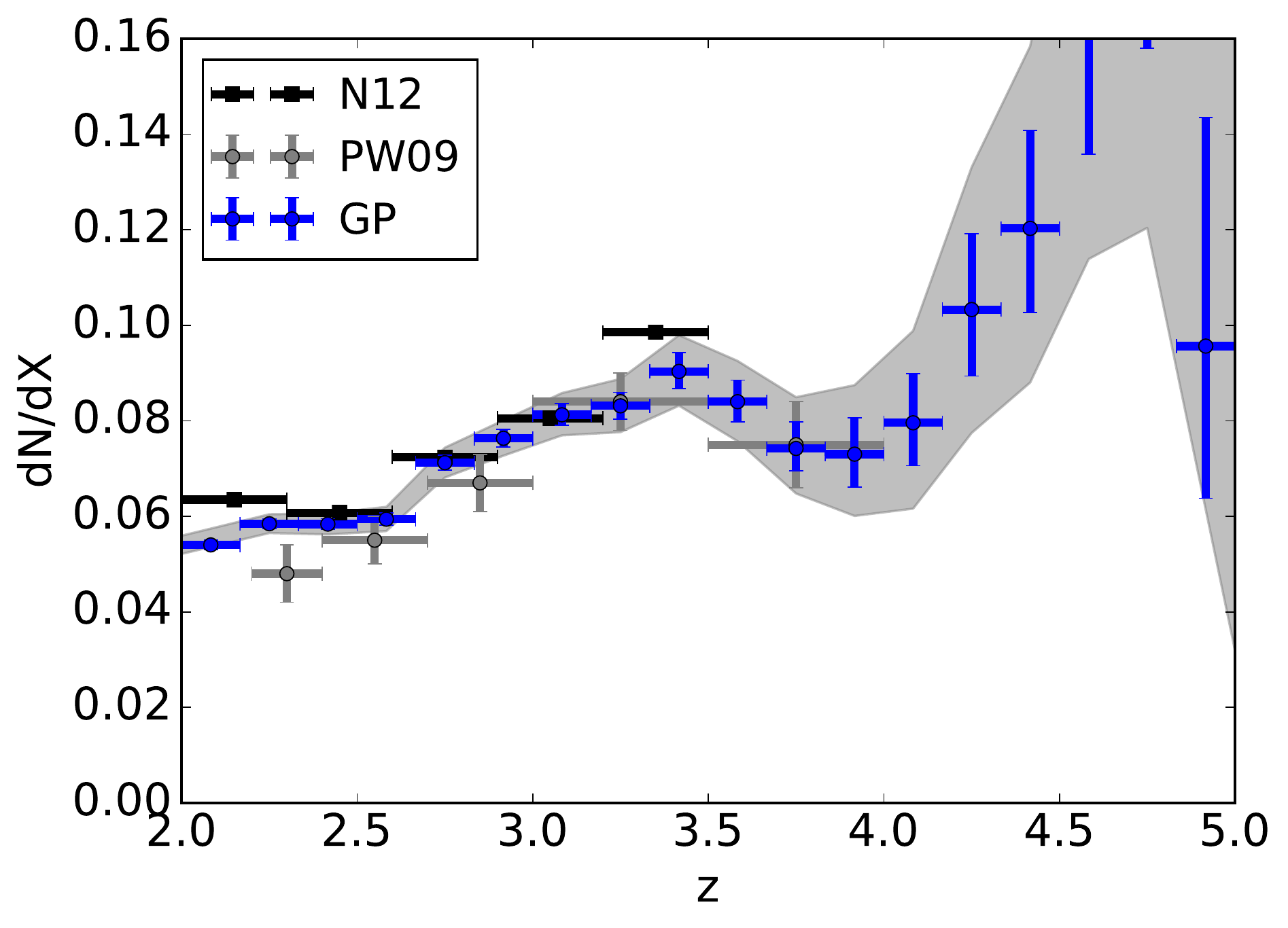}
\caption{
The line density of DLAs as a function of redshift from our DR12 DLA catalogue (blue, `GP'), compared to the results of \protect\cite{Noterdaeme:2012} (N12; black) and \protect\cite{Prochaska:2009} (PW09; grey).
Note that statistical error was not computed in \protect\cite{Noterdaeme:2012}.}
\label{fig:dndx}
\end{figure}

Figure \ref{fig:omega_dla} shows $\Omega_\mathrm{DLA}$, the HI density in DLAs in units of the cosmic density (see also Table \ref{tab:DR12/dndx_all.txt}). Our results are consistent with those of
\cite{Noterdaeme:2012}. As discussed in \cite{Noterdaeme:2012}and \cite{Sanchez:2016}, the discrepancy between \cite{Noterdaeme:2012} and \cite{Prochaska:2009} is driven by the relatively small 
size of the SDSS DR5 dataset used in \cite{Prochaska:2009}. In DR5 there were relatively large statistical errors for $\NHI > 3\times 10^{21}$~\NHunit, 
and a statistical fluctuation caused these bins to be lower than revealed by later datasets.

We also show, for the first time from SDSS, measurements of $\Omega_\mathrm{DLA}$ at $z>4$. As in previous Figures, we
include data only for $z < 5$, as at higher redshifts our $95\%$ confidence limits are consistent with zero DLAs.
Due to the relatively small sample (and noisy spectra) at $z > 4$, we do not convincingly detect DLAs with $\NHI > 3 \times 10^{21}$~\NHunit at this redshift (see Figure~\ref{fig:cddf_zz}).
For column densities $2 \times 10^{20} < \NHI < 10^{21}$, our CDDF is roughly a power law with slope $-1.7$, and for $\NHI > 10^{21}$ it steepens and has a slope roughly $-2.5$.
Thus we a priori expect column densities with $\NHI > 3 \times 10^{21}$~\NHunit\ to contribute around $20\%$ of $\Omega_\mathrm{DLA}$. This expectation is incorporated in the prior CDDF applied to 
spectra where it is not possible to precisely estimate the DLA column density, and implies that a future, larger, sample, should decrease our statistical uncertainty, but not increase 
the measured value of $\Omega_\mathrm{DLA}$.

Figure~\ref{fig:omega_dla}~compares our measurements to those of \cite{Crighton:2015}, who used a smaller survey of high signal to noise spectra.
While our data prefers a slightly higher value of $\Omega_\mathrm{DLA}$ than theirs, this discrepancy is smaller than it appears. Like us, 
\cite{Crighton:2015} securely detect DLAs to $\NHI > 3 \times 10^{21}$~\NHunit, but, when computing $\Omega_\mathrm{DLA}$, they do not include a contribution from DLAs with column densities larger than they detect. If we approximate their method by integrating Eq.~\ref{eq:omdla} only to the largest column density detected in our dataset, our measurement of $\Omega_\mathrm{DLA}$ reduces by about $1.5\,\sigma$ at $z>3.5$, completely reconciling our results.

Figure \ref{fig:dndx} shows the line density of DLAs (see Table \ref{tab:DR12/dndx_all.txt}). Here our results are consistent with those of \cite{Prochaska:2009} and \cite{Noterdaeme:2012} where these two agree. Where they disagree we are between the two at low redshift and consistent with \cite{Prochaska:2009} at high redshift.
$\mathrm{d}N/\mathrm{d}X$ is sensitive to the weakest DLAs, making some earlier methods prone to false positives, especially at high redshifts where there is strong absorption from the \Lya~forest.
Indeed, \cite{Noterdaeme:2012} perform a substantial correction to their highest redshift bin using mocks, and our results suggest this correction was moderately too small.
Our method is robust to this problem; in the presence of strong absorption from the forest it simply reduces the strength of the detection.
We see that the template fitting of \cite{Noterdaeme:2012} also overestimates $\mathrm{d}N/\mathrm{d}X$ in the lowest redshift bin. This may be due to the difficulty of securely detecting DLAs
using templates when the spectrum is too short to include the full damping wing.

Both $\Omega_\mathrm{DLA}$ and $\mathrm{d}N/\mathrm{d}X$ exhibit a similar pattern of evolution with redshift. From $z=2$\,--\,$2.75$ they evolve minimally, if at all. Between $z=2.75$ and $z=3.5$ they both increase gently to higher redshifts.
This increase is now detected at fairly high significance, and is a little curious as it does not appear to be reproduced in cosmological
simulations \citep[e.g.][]{Bird:2014, Rahmati:2015}.

A constant $\Omega_\mathrm{DLA}$ at $z>3.5$ lies close to our 95\% confidence limits, suggesting that what evolution appears may be statistical. 
However, $\mathrm{d}N/\mathrm{d}X$ shows a noticeable dip between $z=3.5$ and $z=4.5$, and
an increase for $z>4.5$. We extracted $\mathrm{d}N/\mathrm{d}X$ from the simulations 
used in \cite{Bird:2014}, and found that it was roughly constant at these redshifts. 
One possibility is that at these high redshifts, where strong \Lya forest absorption makes estimating the neutral hydrogen column 
density more difficult, we are mistaking lower column density systems for DLAs. Another is that this is simply a rare statistical fluctuation.

One possible explanation is selection effects from SDSS colour selection.
The quasar rest-frame Lyman limit at $912$~\AA~moves into the observable region of the spectrum at $z=3.7$.
\cite{Prochaska:2009a} identified a systematic in the SDSS colour selection which causes quasar sightlines containing Lyman Limit Systems (LLS) to
be preferentially selected for spectroscopic followup. \cite{Worseck:2011} and \cite{Fumagalli:2013} showed that, due to the width 
of the $u$-band filter in SDSS,  LLS are over-sampled for $z=2.5-3.6$ for all quasars in the redshift range $z=3.0-3.6$.
In order to determine whether this systematic was affecting our results, we considered a modified catalogue with all affected spectral 
path excised. We found that the feature at $z=3.5-4.5$ is also present in $z>3.6$ quasars, which do not suffer from this bias. Thus the 
feature persists in a clean sample, albeit at reduced significance. LLS are more far common than DLAs and so the correlation between 
them is weak enough that a biased selection of LLS apparently does not bias the abundance of DLAs. We note however that other, as yet
undiscovered, systematics due to colour selection remain possible.

It is interesting that, although they note that their measurement of the line density has significant
systematic uncertainty, \cite{Crighton:2015} also observe an increase in the line density of DLAs between $z=4$ and $z=4.9$, 
to $\mathrm{d}N/\mathrm{d}X \sim 0.1$, in agreement with our results.
It will be interesting to see whether this trend remains in future datasets.

\section{Checks For Systematics}
\label{sec:systematic}

\subsection{Effect of SNR}
\label{sec:snr}

\begin{figure*}
\includegraphics[width=0.45\textwidth]{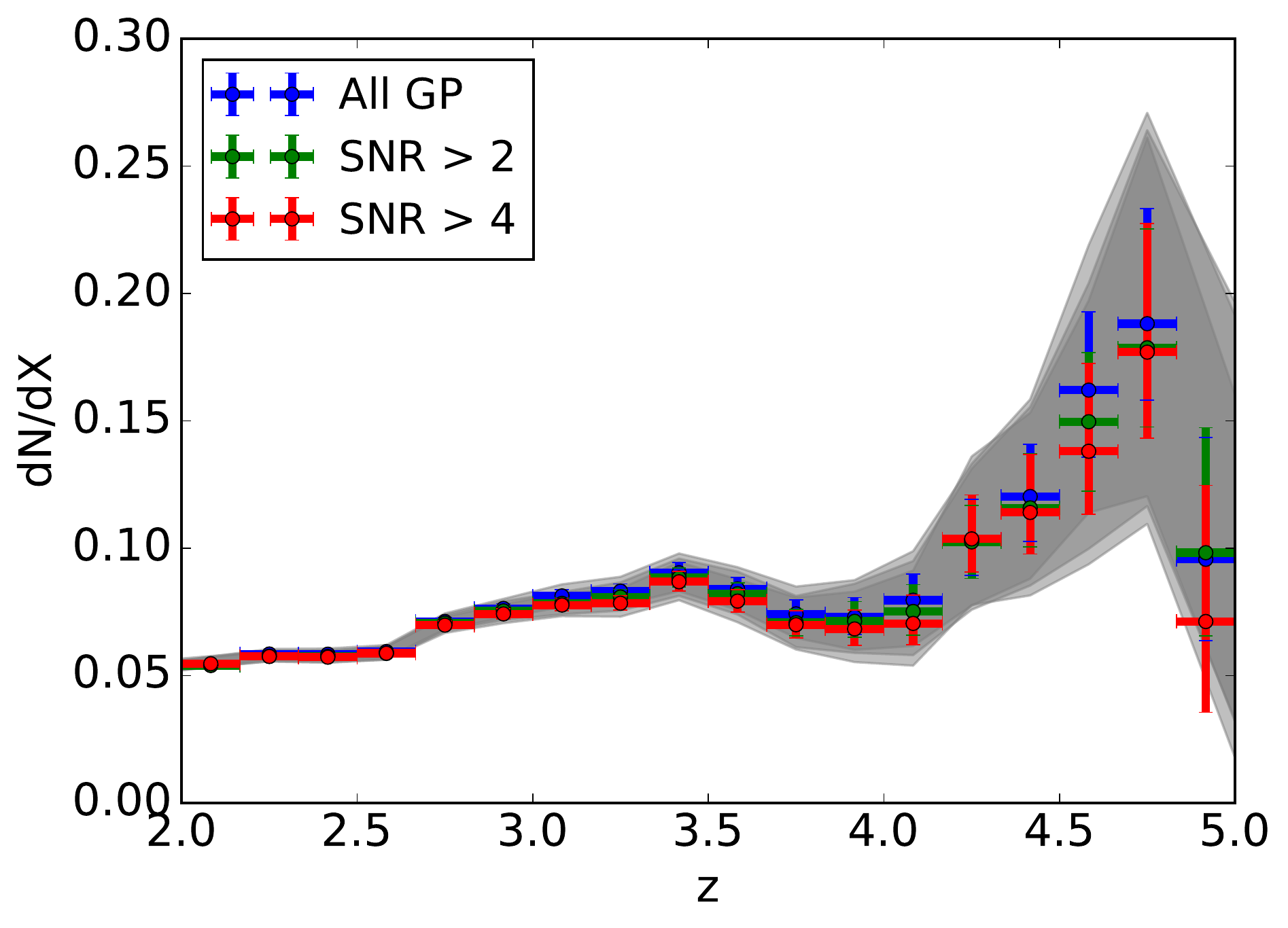}
\includegraphics[width=0.45\textwidth]{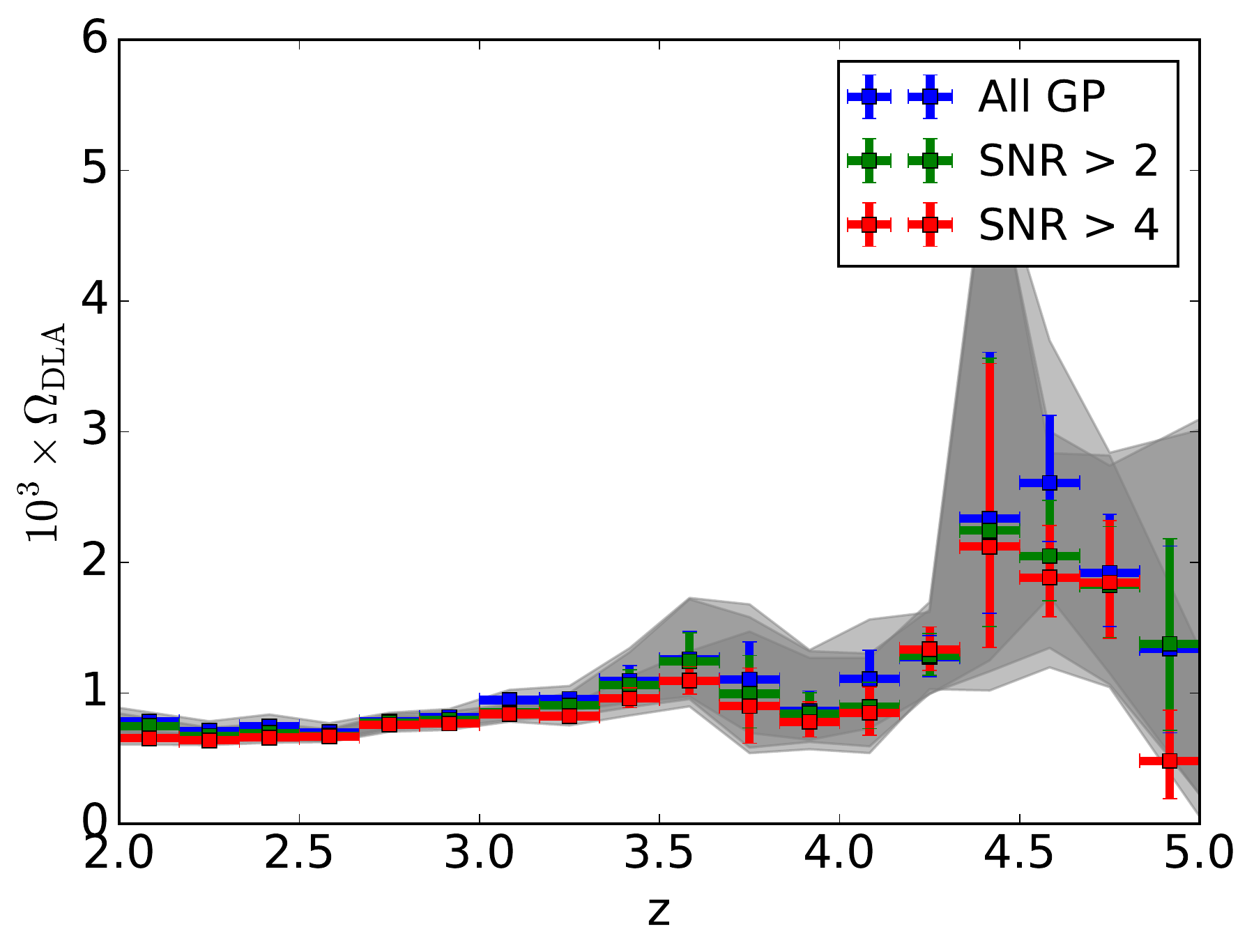}
\caption{
The line density and $\Omega_\mathrm{DLA}$ in DLAs as a function of redshift for a sample with different minimal signal to noise ratios.
The green points show SNR $>2$, which exclude $2\%$ of the noisiest spectra. The red points show SNR $>4$, which
exclude $7.5\%$ of the noisiest spectra. Grey bands show the $95\%$ confidence intervals, which overlap very closely for all samples.
}
\label{fig:snr}
\end{figure*}

Figure \ref{fig:snr} shows the DLA abundance from subsets of the DR12 quasar catalogue with
signal to noise (SNR) $>2$ and $>4$. This demonstrates that our results are insensitive to the SNRs of the spectra.
We define the SNR as the median of the ratio between the flux
and the pixel noise within spectrum redwards
of the rest-frame \Lya emission. We do not use the \Lya~part of the spectrum, to avoid inducing correlations between SNR 
and the presence of a DLA. Since our definition uses a part of the spectrum in which there is strong \Lya emission
and little absorption, it produces a higher SNR than the usual definition using the noise in the forest. Under our definition,
$70$\% of spectra have SNR $> 8$, $92.5$ \% SNR $>4$, and 98 \% SNR $>2$.
The effect of restricting to high SNR spectra is extremely small, less than one half the $68\%$ confidence limits of the measurement, verifying that our results are not biased by noise.

We have also verified explicitly that our results are not sensitive to the noise in a given pixel. To do this, we omitted 
from our catalogue spectral pixels where the signal to noise in the pixel was less than $2$; the derived values for $\Omega_\mathrm{DLA}$ and $\mathrm{d}N/\mathrm{d}X$ changed by less than the $68\%$ confidence limits.

\subsection{Quasar Redshift}
\label{sec:bright}

\begin{figure*}
\includegraphics[width=0.45\textwidth]{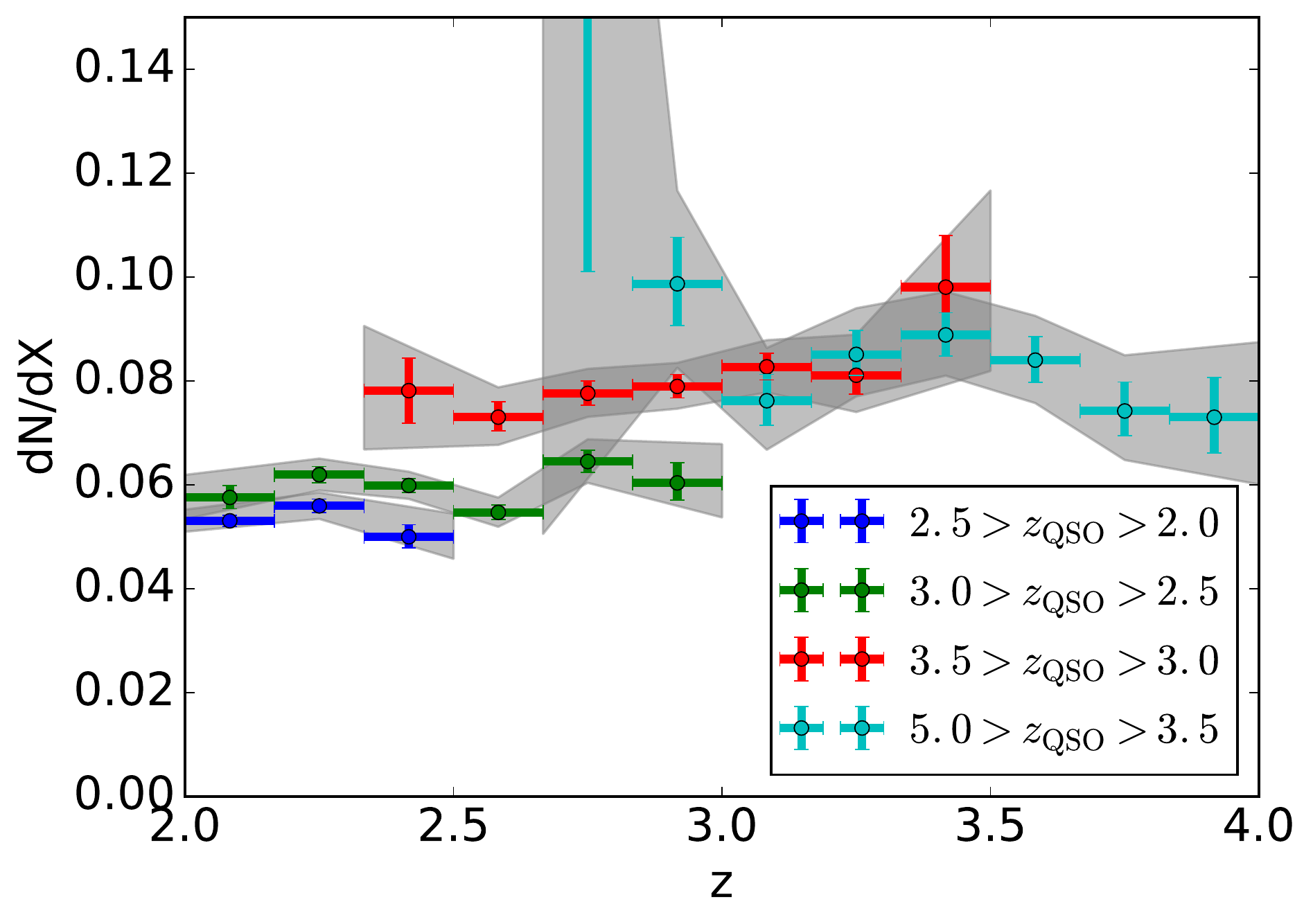}
\includegraphics[width=0.45\textwidth]{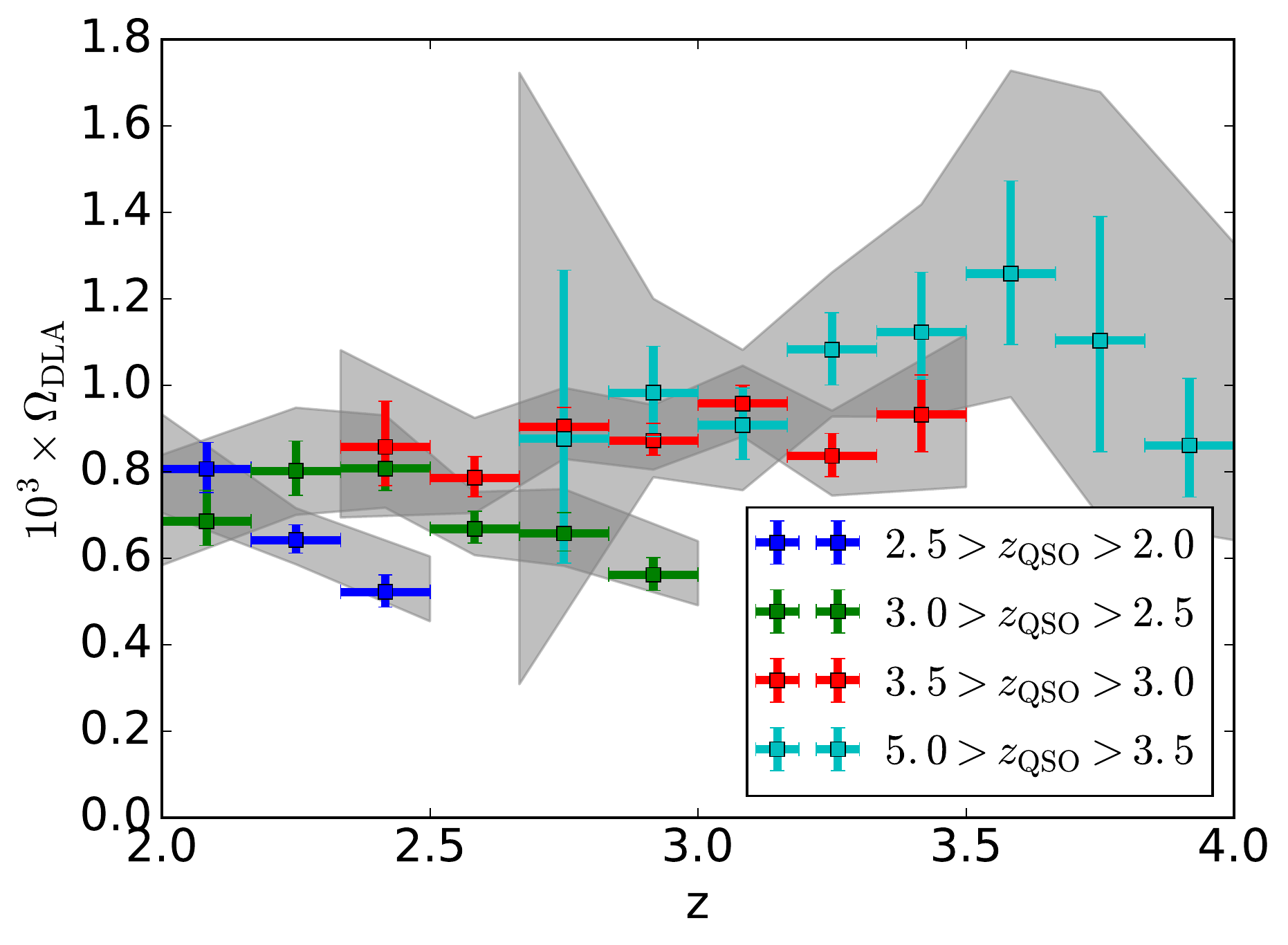}
\caption{
Redshift evolution of the line density in DLAs (Left) and matter density in DLAs, $\Omega_\mathrm{DLA}$ (Right),
as a function of the redshift of the quasar ($z_\mathrm{QSO}$). Correlation between the properties of the background quasar and
those of the absorber may indicate systematic error.
}
\label{fig:zqsos}
\end{figure*}

Due to the large distance separating absorbers from the background quasar, the properties of the absorber should be largely uncorrelated with the properties of the background quasar illuminating it.
However, many common sources of systematic error will occur in the rest frame of the quasar. Figure~\ref{fig:zqsos} shows both $\mathrm{d}N/\mathrm{d}X$ and $\Omega_\mathrm{DLA}$ for our dataset binned as a function of quasar redshift.
In the absence of systematics, we would expect both quantities to be the same irrespective of the redshift of the quasar, and thus the two lines should overlap. This is an extremely powerful test; for example,
if the Lyman-$\beta$ forest part of the spectrum is not removed, the effect of Lyman-$\beta$ absorption is highly visible.

As shown by Figure \ref{fig:zqsos}, there is some residual correlation between DLA properties and quasar redshift, in particular in the line density of the DLAs at $z < 3$.
The majority of this correlation comes from the area nearest the quasar. We found that removing the spectrum with $z - z_\mathrm{QSO} < 0.1$ was sufficient to remove all correlation
between the quasar redshift and $\Omega_\mathrm{DLA}$. Some correlation still remained between $\mathrm{d}N/\mathrm{d}X$ and $z_\mathrm{QSO}$ at $z<2.5$. At $z=3$, $\Delta z = 0.1$ corresponds to
a physical distance of $\sim 1$ Gpc, much larger than the expected mean free path to ionizing photons, $\sim 100$ Mpc \citep[e.g.][]{Haardt:2012}. This signal is thus
unlikely to be an extended proximity zone around quasars. One possibility is some systematic in the data; perhaps a small subset of quasars in this dataset has
sufficient emission near the redshift of the quasar that our Gaussian process code fails to detect the DLA. If such objects were sufficiently rare, they might not
appear in the training set. However, it seems unlikely that an effect of this magnitude could be missed from the training set. Another possibility is some interaction between DLAs and the SDSS colour selection algorithm which mildly biases spectroscopic followup towards quasars
which do not contain a DLA near the redshift of the quasar\footnote{Note that the colour selection bias towards LLS described in \cite{Worseck:2011} does not affect absorbers at $z<2.5$ and so cannot provide a full explanation.}, or a systematic bias in quasar redshift. 

There are also potential explanations which do not involve observational systematics.
For example, quasars are clustered, inhabiting overdense regions. Thus we may be detecting some correlation with the gas in these overdense regions. This could be either a reduction
in gas corresponding to the more advanced evolution of structure growth in the overdense region, or overlapping proximity regions of nearby faint or obscured quasars.

We have investigated removing the region near the quasar redshift, and found that our overall results change by about the $68\%$ confidence limits. As it is unclear
whether the explanation for this discrepancy is observational or physical, we choose to estimate DLA properties using all available data and caution the reader that there may
be unmodelled systematics of the approximate magnitude of the $68\%$ confidence limits.

\section{Conclusion}
\label{sec:conclusions}

We have presented a new measurement of the abundance of neutral hydrogen from $z=2$ to $z=5$, using a catalogue built from SDSS DR12 spectra \citep{Garnett:2016}.
Our results are in good agreement with previous measurements by \cite{Noterdaeme:2012}, \cite{Prochaska:2009} and \cite{Crighton:2015}. The robust statistical 
properties of our method allow us to include partial detections of DLAs in noisy spectra, which allows us to use even the noisiest spectra and 
thus substantially increase the size of the sample. We note, however, that there is a residual dependence on quasar redshift for $z<2.5$, which may indicate 
un-modelled systematic errors within our data, of similar magnitude to the statistical uncertainty.
Our larger sample size allows us to substantially reduce statistical error and extend our measurement to $z > 4$ for 
the first time using SDSS spectra. We detect no evolution in the shape of the CDDF for $2 < z < 4$, while the abundance 
of both DLAs and neutral hydrogen increases moderately over the same redshift range. 

\section*{Acknowledgements}

SB was supported by a McWilliams Fellowship from Carnegie Mellon University and
by NASA through Einstein Postdoctoral Fellowship Award Number PF5-160133.
SH was supported by NASA and NSF for this work.

%

\appendix

\section{Tables of Results}

\begin{table*}
 \centering
\begin{tabular}{cccccc}
\hline
$\log_{10} \mathrm{N}_\mathrm{HI}$ & $f(N_\mathrm{HI})$  $( 10^{ -21 } )$ & $68$\% limits $( 10^{ -21 } )$ & $95$\% limits $( 10^{ -21 } )$ \\
 \hline
$ 20.0 - 20.1 $ & $ 2.392 $ & $ 2.377 - 2.408 $ & $ 2.362 - 2.423 $  \\
$ 20.1 - 20.2 $ & $ 0.538 $ & $ 0.531 - 0.546 $ & $ 0.523 - 0.554 $  \\
$ 20.2 - 20.3 $ & $ 0.300 $ & $ 0.295 - 0.305 $ & $ 0.290 - 0.310 $  \\
$ 20.3 - 20.4 $ & $ 0.200 $ & $ 0.197 - 0.204 $ & $ 0.193 - 0.208 $  \\
$ 20.4 - 20.5 $ & $ 0.141 $ & $ 0.138 - 0.143 $ & $ 0.135 - 0.146 $  \\
$ 20.5 - 20.6 $ & $ 9.84 \times 10^{ -2 }$ & $ [9.64  - 10.05 ]\times 10^{ -2 }$ & $ [9.45  - 10.24 ]\times 10^{ -2 }$  \\
$ 20.6 - 20.7 $ & $ 6.60 \times 10^{ -2 }$ & $ [6.46  - 6.75 ]\times 10^{ -2 }$ & $ [6.32  - 6.89 ]\times 10^{ -2 }$  \\
$ 20.7 - 20.8 $ & $ 4.62 \times 10^{ -2 }$ & $ [4.52  - 4.73 ]\times 10^{ -2 }$ & $ [4.41  - 4.84 ]\times 10^{ -2 }$  \\
$ 20.8 - 20.9 $ & $ 3.13 \times 10^{ -2 }$ & $ [3.05  - 3.21 ]\times 10^{ -2 }$ & $ [2.98  - 3.28 ]\times 10^{ -2 }$  \\
$ 20.9 - 21.0 $ & $ 1.90 \times 10^{ -2 }$ & $ [1.85  - 1.96 ]\times 10^{ -2 }$ & $ [1.80  - 2.01 ]\times 10^{ -2 }$  \\
$ 21.0 - 21.1 $ & $ 1.29 \times 10^{ -2 }$ & $ [1.25  - 1.33 ]\times 10^{ -2 }$ & $ [1.21  - 1.36 ]\times 10^{ -2 }$  \\
$ 21.1 - 21.2 $ & $ 7.79 \times 10^{ -3 }$ & $ [7.52  - 8.06 ]\times 10^{ -3 }$ & $ [7.27  - 8.32 ]\times 10^{ -3 }$  \\
$ 21.2 - 21.3 $ & $ 5.08 \times 10^{ -3 }$ & $ [4.90  - 5.27 ]\times 10^{ -3 }$ & $ [4.72  - 5.45 ]\times 10^{ -3 }$  \\
$ 21.3 - 21.4 $ & $ 2.83 \times 10^{ -3 }$ & $ [2.71  - 2.96 ]\times 10^{ -3 }$ & $ [2.60  - 3.08 ]\times 10^{ -3 }$  \\
$ 21.4 - 21.5 $ & $ 1.75 \times 10^{ -3 }$ & $ [1.67  - 1.83 ]\times 10^{ -3 }$ & $ [1.59  - 1.91 ]\times 10^{ -3 }$  \\
$ 21.5 - 21.6 $ & $ 9.70 \times 10^{ -4 }$ & $ [9.20  - 10.29 ]\times 10^{ -4 }$ & $ [8.70  - 10.78 ]\times 10^{ -4 }$  \\
$ 21.6 - 21.7 $ & $ 5.40 \times 10^{ -4 }$ & $ [5.08  - 5.79 ]\times 10^{ -4 }$ & $ [4.75  - 6.15 ]\times 10^{ -4 }$  \\
$ 21.7 - 21.8 $ & $ 2.80 \times 10^{ -4 }$ & $ [2.57  - 3.06 ]\times 10^{ -4 }$ & $ [2.34  - 3.29 ]\times 10^{ -4 }$  \\
$ 21.8 - 21.9 $ & $ 1.54 \times 10^{ -4 }$ & $ [1.39  - 1.70 ]\times 10^{ -4 }$ & $ [1.27  - 1.86 ]\times 10^{ -4 }$  \\
$ 21.9 - 22.0 $ & $ 7.76 \times 10^{ -5 }$ & $ [6.67  - 8.84 ]\times 10^{ -5 }$ & $ [5.95  - 9.74 ]\times 10^{ -5 }$  \\
$ 22.0 - 22.1 $ & $ 3.44 \times 10^{ -5 }$ & $ [2.87  - 4.16 ]\times 10^{ -5 }$ & $ [2.29  - 4.73 ]\times 10^{ -5 }$  \\
$ 22.1 - 22.2 $ & $ 1.82 \times 10^{ -5 }$ & $ [1.48  - 2.39 ]\times 10^{ -5 }$ & $ [1.14  - 2.85 ]\times 10^{ -5 }$  \\
$ 22.2 - 22.3 $ & $ 1.18 \times 10^{ -5 }$ & $ [9.04  - 16.27 ]\times 10^{ -6 }$ & $ [7.23  - 18.99 ]\times 10^{ -6 }$  \\
$ 22.3 - 22.4 $ & $ 5.75 \times 10^{ -6 }$ & $ [3.59  - 7.90 ]\times 10^{ -6 }$ & $ [2.15  - 10.05 ]\times 10^{ -6 }$  \\
$ 22.4 - 22.5 $ & $ 3.99 \times 10^{ -6 }$ & $ [2.85  - 6.27 ]\times 10^{ -6 }$ & $ [1.71  - 7.99 ]\times 10^{ -6 }$  \\
$ 22.5 - 22.6 $ & $ 2.27 \times 10^{ -6 }$ & $ [1.36  - 3.63 ]\times 10^{ -6 }$ & $ [4.53  - 49.84 ]\times 10^{ -7 }$  \\
$ 22.6 - 22.7 $ & $ 1.80 \times 10^{ -6 }$ & $ [1.08  - 2.88 ]\times 10^{ -6 }$ & $ [3.60  - 39.59 ]\times 10^{ -7 }$  \\
$ 22.7 - 22.8 $ & $ 1.14 \times 10^{ -6 }$ & $ [5.72  - 20.01 ]\times 10^{ -7 }$ & $ [2.86  - 28.59 ]\times 10^{ -7 }$  \\
$ 22.8 - 22.9 $ & $ 6.81 \times 10^{ -7 }$ & $ [4.54  - 13.63 ]\times 10^{ -7 }$ & $0 -  1.82 \times 10^{ -6 }$  \\
$ 22.9 - 23.0 $ & $ 3.61 \times 10^{ -7 }$ & $ [1.80  - 9.02 ]\times 10^{ -7 }$ & $0 -  1.26 \times 10^{ -6 }$  \\
\hline
  \end{tabular}
 \caption{Average column density distribution function for all DLAs with $2<z<5$. See also Figure \ref{fig:cddf}.}
\label{tab:DR12/cddf_all.txt}
 \end{table*}

\begin{table*}
 \centering
\begin{tabular}{cccccc}
\hline
$\log_{10} \mathrm{N}_\mathrm{HI}$ & $f(N_\mathrm{HI})$  $( 10^{ -21 } )$ & $68$\% limits $( 10^{ -21 } )$ & $95$\% limits $( 10^{ -21 } )$ \\
 \hline
$ 20.0 - 20.1 $ & $ 2.112 $ & $ 2.091 - 2.132 $ & $ 2.072 - 2.152 $  \\
$ 20.1 - 20.2 $ & $ 0.498 $ & $ 0.488 - 0.508 $ & $ 0.478 - 0.518 $  \\
$ 20.2 - 20.3 $ & $ 0.277 $ & $ 0.271 - 0.284 $ & $ 0.264 - 0.291 $  \\
$ 20.3 - 20.4 $ & $ 0.181 $ & $ 0.177 - 0.186 $ & $ 0.172 - 0.191 $  \\
$ 20.4 - 20.5 $ & $ 0.129 $ & $ 0.125 - 0.132 $ & $ 0.122 - 0.136 $  \\
$ 20.5 - 20.6 $ & $ 9.10 \times 10^{ -2 }$ & $ [8.84  - 9.38 ]\times 10^{ -2 }$ & $ [8.59  - 9.63 ]\times 10^{ -2 }$  \\
$ 20.6 - 20.7 $ & $ 6.07 \times 10^{ -2 }$ & $ [5.88  - 6.27 ]\times 10^{ -2 }$ & $ [5.70  - 6.46 ]\times 10^{ -2 }$  \\
$ 20.7 - 20.8 $ & $ 4.12 \times 10^{ -2 }$ & $ [3.98  - 4.27 ]\times 10^{ -2 }$ & $ [3.85  - 4.41 ]\times 10^{ -2 }$  \\
$ 20.8 - 20.9 $ & $ 2.77 \times 10^{ -2 }$ & $ [2.67  - 2.88 ]\times 10^{ -2 }$ & $ [2.58  - 2.98 ]\times 10^{ -2 }$  \\
$ 20.9 - 21.0 $ & $ 1.72 \times 10^{ -2 }$ & $ [1.65  - 1.80 ]\times 10^{ -2 }$ & $ [1.58  - 1.86 ]\times 10^{ -2 }$  \\
$ 21.0 - 21.1 $ & $ 1.15 \times 10^{ -2 }$ & $ [1.10  - 1.20 ]\times 10^{ -2 }$ & $ [1.05  - 1.25 ]\times 10^{ -2 }$  \\
$ 21.1 - 21.2 $ & $ 7.27 \times 10^{ -3 }$ & $ [6.93  - 7.64 ]\times 10^{ -3 }$ & $ [6.58  - 8.01 ]\times 10^{ -3 }$  \\
$ 21.2 - 21.3 $ & $ 4.56 \times 10^{ -3 }$ & $ [4.31  - 4.82 ]\times 10^{ -3 }$ & $ [4.07  - 5.06 ]\times 10^{ -3 }$  \\
$ 21.3 - 21.4 $ & $ 2.56 \times 10^{ -3 }$ & $ [2.39  - 2.73 ]\times 10^{ -3 }$ & $ [2.24  - 2.90 ]\times 10^{ -3 }$  \\
$ 21.4 - 21.5 $ & $ 1.67 \times 10^{ -3 }$ & $ [1.56  - 1.80 ]\times 10^{ -3 }$ & $ [1.46  - 1.91 ]\times 10^{ -3 }$  \\
$ 21.5 - 21.6 $ & $ 8.56 \times 10^{ -4 }$ & $ [7.91  - 9.38 ]\times 10^{ -4 }$ & $ [7.26  - 10.11 ]\times 10^{ -4 }$  \\
$ 21.6 - 21.7 $ & $ 4.79 \times 10^{ -4 }$ & $ [4.34  - 5.31 ]\times 10^{ -4 }$ & $ [3.89  - 5.77 ]\times 10^{ -4 }$  \\
$ 21.7 - 21.8 $ & $ 2.73 \times 10^{ -4 }$ & $ [2.42  - 3.09 ]\times 10^{ -4 }$ & $ [2.16  - 3.40 ]\times 10^{ -4 }$  \\
$ 21.8 - 21.9 $ & $ 1.39 \times 10^{ -4 }$ & $ [1.23  - 1.64 ]\times 10^{ -4 }$ & $ [1.02  - 1.88 ]\times 10^{ -4 }$  \\
$ 21.9 - 22.0 $ & $ 7.79 \times 10^{ -5 }$ & $ [6.49  - 9.74 ]\times 10^{ -5 }$ & $ [5.52  - 11.04 ]\times 10^{ -5 }$  \\
$ 22.0 - 22.1 $ & $ 3.87 \times 10^{ -5 }$ & $ [3.10  - 4.90 ]\times 10^{ -5 }$ & $ [2.32  - 5.93 ]\times 10^{ -5 }$  \\
$ 22.1 - 22.2 $ & $ 2.05 \times 10^{ -5 }$ & $ [1.43  - 2.87 ]\times 10^{ -5 }$ & $ [1.02  - 3.48 ]\times 10^{ -5 }$  \\
$ 22.2 - 22.3 $ & $ 9.77 \times 10^{ -6 }$ & $ [6.51  - 16.28 ]\times 10^{ -6 }$ & $ [3.26  - 19.53 ]\times 10^{ -6 }$  \\
$ 22.3 - 22.4 $ & $ 5.17 \times 10^{ -6 }$ & $ [2.59  - 9.05 ]\times 10^{ -6 }$ & $ [1.29  - 11.63 ]\times 10^{ -6 }$  \\
$ 22.4 - 22.5 $ & $ 4.11 \times 10^{ -6 }$ & $ [2.05  - 7.19 ]\times 10^{ -6 }$ & $ [1.03  - 9.24 ]\times 10^{ -6 }$  \\
$ 22.5 - 22.6 $ & $ 2.45 \times 10^{ -6 }$ & $ [1.63  - 4.89 ]\times 10^{ -6 }$ & $0 -  6.53 \times 10^{ -6 }$  \\
$ 22.6 - 22.7 $ & $ 1.94 \times 10^{ -6 }$ & $ [1.30  - 3.89 ]\times 10^{ -6 }$ & $0 -  5.18 \times 10^{ -6 }$  \\
$ 22.7 - 22.8 $ & $ 1.54 \times 10^{ -6 }$ & $ [5.15  - 30.88 ]\times 10^{ -7 }$ & $0 -  4.12 \times 10^{ -6 }$  \\
$ 22.8 - 22.9 $ & $ 8.18 \times 10^{ -7 }$ & $ [4.09  - 20.44 ]\times 10^{ -7 }$ & $0 -  2.86 \times 10^{ -6 }$  \\
$ 22.9 - 23.0 $ & $ 3.25 \times 10^{ -7 }$ & $0 -  1.30 \times 10^{ -6 }$ & $0 -  1.62 \times 10^{ -6 }$  \\
\hline
  \end{tabular}
 \caption{Column density distribution function for all DLAs with $2<z<2.5$. See also Figure \ref{fig:cddf_zz}.}
\label{tab:DR12/cddf_z225.txt}
 \end{table*}

\begin{table*}
 \centering
\begin{tabular}{cccccc}
\hline
$\log_{10} \mathrm{N}_\mathrm{HI}$ & $f(N_\mathrm{HI})$  $( 10^{ -21 } )$ & $68$\% limits $( 10^{ -21 } )$ & $95$\% limits $( 10^{ -21 } )$ \\
 \hline
$ 20.0 - 20.1 $ & $ 2.824 $ & $ 2.793 - 2.856 $ & $ 2.763 - 2.887 $  \\
$ 20.1 - 20.2 $ & $ 0.570 $ & $ 0.554 - 0.586 $ & $ 0.540 - 0.600 $  \\
$ 20.2 - 20.3 $ & $ 0.324 $ & $ 0.314 - 0.335 $ & $ 0.305 - 0.345 $  \\
$ 20.3 - 20.4 $ & $ 0.222 $ & $ 0.215 - 0.230 $ & $ 0.208 - 0.237 $  \\
$ 20.4 - 20.5 $ & $ 0.149 $ & $ 0.144 - 0.155 $ & $ 0.139 - 0.160 $  \\
$ 20.5 - 20.6 $ & $ 0.108 $ & $ 0.104 - 0.112 $ & $ [9.99  - 11.59 ]\times 10^{ -2 }$  \\
$ 20.6 - 20.7 $ & $ 7.01 \times 10^{ -2 }$ & $ [6.72  - 7.31 ]\times 10^{ -2 }$ & $ [6.45  - 7.59 ]\times 10^{ -2 }$  \\
$ 20.7 - 20.8 $ & $ 5.12 \times 10^{ -2 }$ & $ [4.90  - 5.34 ]\times 10^{ -2 }$ & $ [4.70  - 5.55 ]\times 10^{ -2 }$  \\
$ 20.8 - 20.9 $ & $ 3.51 \times 10^{ -2 }$ & $ [3.35  - 3.67 ]\times 10^{ -2 }$ & $ [3.21  - 3.82 ]\times 10^{ -2 }$  \\
$ 20.9 - 21.0 $ & $ 2.08 \times 10^{ -2 }$ & $ [1.98  - 2.19 ]\times 10^{ -2 }$ & $ [1.88  - 2.30 ]\times 10^{ -2 }$  \\
$ 21.0 - 21.1 $ & $ 1.39 \times 10^{ -2 }$ & $ [1.31  - 1.46 ]\times 10^{ -2 }$ & $ [1.24  - 1.54 ]\times 10^{ -2 }$  \\
$ 21.1 - 21.2 $ & $ 8.38 \times 10^{ -3 }$ & $ [7.86  - 8.94 ]\times 10^{ -3 }$ & $ [7.38  - 9.46 ]\times 10^{ -3 }$  \\
$ 21.2 - 21.3 $ & $ 5.45 \times 10^{ -3 }$ & $ [5.11  - 5.83 ]\times 10^{ -3 }$ & $ [4.80  - 6.21 ]\times 10^{ -3 }$  \\
$ 21.3 - 21.4 $ & $ 3.00 \times 10^{ -3 }$ & $ [2.78  - 3.27 ]\times 10^{ -3 }$ & $ [2.56  - 3.49 ]\times 10^{ -3 }$  \\
$ 21.4 - 21.5 $ & $ 1.82 \times 10^{ -3 }$ & $ [1.67  - 1.99 ]\times 10^{ -3 }$ & $ [1.54  - 2.14 ]\times 10^{ -3 }$  \\
$ 21.5 - 21.6 $ & $ 1.03 \times 10^{ -3 }$ & $ [9.28  - 11.35 ]\times 10^{ -4 }$ & $ [8.42  - 12.38 ]\times 10^{ -4 }$  \\
$ 21.6 - 21.7 $ & $ 5.46 \times 10^{ -4 }$ & $ [4.78  - 6.14 ]\times 10^{ -4 }$ & $ [4.23  - 6.83 ]\times 10^{ -4 }$  \\
$ 21.7 - 21.8 $ & $ 2.49 \times 10^{ -4 }$ & $ [2.17  - 3.04 ]\times 10^{ -4 }$ & $ [1.74  - 3.47 ]\times 10^{ -4 }$  \\
$ 21.8 - 21.9 $ & $ 1.38 \times 10^{ -4 }$ & $ [1.12  - 1.72 ]\times 10^{ -4 }$ & $ [9.48  - 19.82 ]\times 10^{ -5 }$  \\
$ 21.9 - 22.0 $ & $ 5.48 \times 10^{ -5 }$ & $ [4.11  - 7.53 ]\times 10^{ -5 }$ & $ [3.42  - 9.58 ]\times 10^{ -5 }$  \\
$ 22.0 - 22.1 $ & $ 1.09 \times 10^{ -5 }$ & $ [5.44  - 21.75 ]\times 10^{ -6 }$ & $ [5.44  - 32.62 ]\times 10^{ -6 }$  \\
$ 22.1 - 22.2 $ & $ 8.64 \times 10^{ -6 }$ & $ [4.32  - 21.59 ]\times 10^{ -6 }$ & $0 -  2.59 \times 10^{ -5 }$  \\
$ 22.2 - 22.3 $ & $ 1.03 \times 10^{ -5 }$ & $ [6.86  - 17.15 ]\times 10^{ -6 }$ & $ [3.43  - 20.58 ]\times 10^{ -6 }$  \\
$ 22.3 - 22.4 $ & $ 2.72 \times 10^{ -6 }$ & $0 -  8.17 \times 10^{ -6 }$ & $0 -  1.36 \times 10^{ -5 }$  \\
$ 22.4 - 22.5 $ & $ 2.16 \times 10^{ -6 }$ & $0 -  6.49 \times 10^{ -6 }$ & $0 -  8.66 \times 10^{ -6 }$  \\
$ 22.5 - 22.6 $ & $0$ & $0 -  3.44 \times 10^{ -6 }$ & $0 -  5.16 \times 10^{ -6 }$  \\
$ 22.6 - 22.7 $ & $0$ & $0 -  2.73 \times 10^{ -6 }$ & $0 -  2.73 \times 10^{ -6 }$  \\
$ 22.7 - 22.8 $ & $0$ & $0 -  1.08 \times 10^{ -6 }$ & $0 -  2.17 \times 10^{ -6 }$  \\
$ 22.8 - 22.9 $ & $0$ & $0 -  8.62 \times 10^{ -7 }$ & $0 -  1.72 \times 10^{ -6 }$  \\
$ 22.9 - 23.0 $ & $0$ & $0 -  6.84 \times 10^{ -7 }$ & $0 -  1.37 \times 10^{ -6 }$  \\
\hline
  \end{tabular}
 \caption{Column density distribution function for all DLAs with $2.5<z<3$. See also Figure \ref{fig:cddf_zz}.}
\label{tab:DR12/cddf_z253.txt}
 \end{table*}

\begin{table*}
 \centering
\begin{tabular}{cccccc}
\hline
$\log_{10} \mathrm{N}_\mathrm{HI}$ & $f(N_\mathrm{HI})$  $( 10^{ -21 } )$ & $68$\% limits $( 10^{ -21 } )$ & $95$\% limits $( 10^{ -21 } )$ \\
 \hline
$ 20.0 - 20.1 $ & $ 3.288 $ & $ 3.240 - 3.338 $ & $ 3.192 - 3.386 $  \\
$ 20.1 - 20.2 $ & $ 0.715 $ & $ 0.690 - 0.741 $ & $ 0.667 - 0.766 $  \\
$ 20.2 - 20.3 $ & $ 0.381 $ & $ 0.366 - 0.399 $ & $ 0.351 - 0.414 $  \\
$ 20.3 - 20.4 $ & $ 0.261 $ & $ 0.249 - 0.273 $ & $ 0.238 - 0.284 $  \\
$ 20.4 - 20.5 $ & $ 0.192 $ & $ 0.184 - 0.202 $ & $ 0.175 - 0.211 $  \\
$ 20.5 - 20.6 $ & $ 0.123 $ & $ 0.117 - 0.130 $ & $ 0.111 - 0.136 $  \\
$ 20.6 - 20.7 $ & $ 8.83 \times 10^{ -2 }$ & $ [8.35  - 9.31 ]\times 10^{ -2 }$ & $ [7.93  - 9.78 ]\times 10^{ -2 }$  \\
$ 20.7 - 20.8 $ & $ 6.42 \times 10^{ -2 }$ & $ [6.07  - 6.78 ]\times 10^{ -2 }$ & $ [5.73  - 7.13 ]\times 10^{ -2 }$  \\
$ 20.8 - 20.9 $ & $ 4.29 \times 10^{ -2 }$ & $ [4.05  - 4.55 ]\times 10^{ -2 }$ & $ [3.82  - 4.80 ]\times 10^{ -2 }$  \\
$ 20.9 - 21.0 $ & $ 2.47 \times 10^{ -2 }$ & $ [2.30  - 2.65 ]\times 10^{ -2 }$ & $ [2.15  - 2.81 ]\times 10^{ -2 }$  \\
$ 21.0 - 21.1 $ & $ 1.73 \times 10^{ -2 }$ & $ [1.63  - 1.86 ]\times 10^{ -2 }$ & $ [1.53  - 1.97 ]\times 10^{ -2 }$  \\
$ 21.1 - 21.2 $ & $ 9.24 \times 10^{ -3 }$ & $ [8.58  - 10.09 ]\times 10^{ -3 }$ & $ [7.83  - 10.85 ]\times 10^{ -3 }$  \\
$ 21.2 - 21.3 $ & $ 6.89 \times 10^{ -3 }$ & $ [6.44  - 7.49 ]\times 10^{ -3 }$ & $ [5.92  - 8.02 ]\times 10^{ -3 }$  \\
$ 21.3 - 21.4 $ & $ 3.75 \times 10^{ -3 }$ & $ [3.45  - 4.11 ]\times 10^{ -3 }$ & $ [3.15  - 4.46 ]\times 10^{ -3 }$  \\
$ 21.4 - 21.5 $ & $ 2.03 \times 10^{ -3 }$ & $ [1.84  - 2.32 ]\times 10^{ -3 }$ & $ [1.65  - 2.51 ]\times 10^{ -3 }$  \\
$ 21.5 - 21.6 $ & $ 1.35 \times 10^{ -3 }$ & $ [1.20  - 1.54 ]\times 10^{ -3 }$ & $ [1.09  - 1.69 ]\times 10^{ -3 }$  \\
$ 21.6 - 21.7 $ & $ 8.05 \times 10^{ -4 }$ & $ [7.16  - 9.25 ]\times 10^{ -4 }$ & $ [6.26  - 10.44 ]\times 10^{ -4 }$  \\
$ 21.7 - 21.8 $ & $ 3.79 \times 10^{ -4 }$ & $ [3.08  - 4.74 ]\times 10^{ -4 }$ & $ [2.61  - 5.45 ]\times 10^{ -4 }$  \\
$ 21.8 - 21.9 $ & $ 2.45 \times 10^{ -4 }$ & $ [2.07  - 3.01 ]\times 10^{ -4 }$ & $ [1.69  - 3.58 ]\times 10^{ -4 }$  \\
$ 21.9 - 22.0 $ & $ 7.48 \times 10^{ -5 }$ & $ [5.98  - 11.96 ]\times 10^{ -5 }$ & $ [2.99  - 14.95 ]\times 10^{ -5 }$  \\
$ 22.0 - 22.1 $ & $ 2.38 \times 10^{ -5 }$ & $ [1.19  - 4.75 ]\times 10^{ -5 }$ & $0 -  7.13 \times 10^{ -5 }$  \\
$ 22.1 - 22.2 $ & $ 2.83 \times 10^{ -5 }$ & $ [1.89  - 4.72 ]\times 10^{ -5 }$ & $ [9.43  - 66.03 ]\times 10^{ -6 }$  \\
$ 22.2 - 22.3 $ & $ 2.25 \times 10^{ -5 }$ & $ [1.50  - 3.75 ]\times 10^{ -5 }$ & $ [7.49  - 44.96 ]\times 10^{ -6 }$  \\
$ 22.3 - 22.4 $ & $ 5.95 \times 10^{ -6 }$ & $ [5.95  - 17.86 ]\times 10^{ -6 }$ & $0 -  2.98 \times 10^{ -5 }$  \\
$ 22.4 - 22.5 $ & $ 4.73 \times 10^{ -6 }$ & $0 -  1.42 \times 10^{ -5 }$ & $0 -  1.89 \times 10^{ -5 }$  \\
$ 22.5 - 22.6 $ & $ 3.76 \times 10^{ -6 }$ & $0 -  1.13 \times 10^{ -5 }$ & $0 -  1.50 \times 10^{ -5 }$  \\
$ 22.6 - 22.7 $ & $ 2.98 \times 10^{ -6 }$ & $0 -  8.95 \times 10^{ -6 }$ & $0 -  1.19 \times 10^{ -5 }$  \\
$ 22.7 - 22.8 $ & $0$ & $0 -  4.74 \times 10^{ -6 }$ & $0 -  7.11 \times 10^{ -6 }$  \\
$ 22.8 - 22.9 $ & $0$ & $0 -  1.88 \times 10^{ -6 }$ & $0 -  3.76 \times 10^{ -6 }$  \\
$ 22.9 - 23.0 $ & $0$ & $0 -  1.50 \times 10^{ -6 }$ & $0 -  1.50 \times 10^{ -6 }$  \\

\hline
  \end{tabular}
 \caption{Column density distribution function for all DLAs with $3<z<4$. See also Figure \ref{fig:cddf_zz}.}
\label{tab:DR12/cddf_z34.txt}
 \end{table*}

\begin{table*}
 \centering
\begin{tabular}{cccccc}
\hline
$\log_{10} \mathrm{N}_\mathrm{HI}$ & $f(N_\mathrm{HI})$  $( 10^{ -21 } )$ & $68$\% limits $( 10^{ -21 } )$ & $95$\% limits $( 10^{ -21 } )$ \\
 \hline

 $ 20.0 - 20.1 $ & $ 2.735 $ & $ 2.547 - 2.962 $ & $ 2.377 - 3.150 $  \\
$ 20.1 - 20.2 $ & $ 0.839 $ & $ 0.734 - 0.959 $ & $ 0.629 - 1.079 $  \\
$ 20.2 - 20.3 $ & $ 0.500 $ & $ 0.428 - 0.595 $ & $ 0.357 - 0.667 $  \\
$ 20.3 - 20.4 $ & $ 0.284 $ & $ 0.236 - 0.350 $ & $ 0.189 - 0.397 $  \\
$ 20.4 - 20.5 $ & $ 0.203 $ & $ 0.165 - 0.248 $ & $ 0.128 - 0.285 $  \\
$ 20.5 - 20.6 $ & $ 0.155 $ & $ 0.125 - 0.191 $ & $ 0.101 - 0.221 $  \\
$ 20.6 - 20.7 $ & $ 0.118 $ & $ [9.48  - 14.69 ]\times 10^{ -2 }$ & $ [7.58  - 17.06 ]\times 10^{ -2 }$  \\
$ 20.7 - 20.8 $ & $ 7.15 \times 10^{ -2 }$ & $ [5.65  - 9.41 ]\times 10^{ -2 }$ & $ [4.14  - 10.92 ]\times 10^{ -2 }$  \\
$ 20.8 - 20.9 $ & $ 5.68 \times 10^{ -2 }$ & $ [4.48  - 7.18 ]\times 10^{ -2 }$ & $ [3.59  - 8.67 ]\times 10^{ -2 }$  \\
$ 20.9 - 21.0 $ & $ 3.80 \times 10^{ -2 }$ & $ [3.09  - 4.99 ]\times 10^{ -2 }$ & $ [2.14  - 5.94 ]\times 10^{ -2 }$  \\
$ 21.0 - 21.1 $ & $ 2.83 \times 10^{ -2 }$ & $ [2.26  - 3.77 ]\times 10^{ -2 }$ & $ [1.70  - 4.34 ]\times 10^{ -2 }$  \\
$ 21.1 - 21.2 $ & $ 1.65 \times 10^{ -2 }$ & $ [1.20  - 2.25 ]\times 10^{ -2 }$ & $ [8.99  - 26.97 ]\times 10^{ -3 }$  \\
$ 21.2 - 21.3 $ & $ 1.07 \times 10^{ -2 }$ & $ [8.33  - 15.47 ]\times 10^{ -3 }$ & $ [4.76  - 19.04 ]\times 10^{ -3 }$  \\
$ 21.3 - 21.4 $ & $ 7.56 \times 10^{ -3 }$ & $ [4.73  - 10.40 ]\times 10^{ -3 }$ & $ [2.84  - 13.24 ]\times 10^{ -3 }$  \\
$ 21.4 - 21.5 $ & $ 3.76 \times 10^{ -3 }$ & $ [2.25  - 5.26 ]\times 10^{ -3 }$ & $ [1.50  - 6.76 ]\times 10^{ -3 }$  \\
$ 21.5 - 21.6 $ & $ 2.39 \times 10^{ -3 }$ & $ [1.19  - 3.58 ]\times 10^{ -3 }$ & $ [5.97  - 47.72 ]\times 10^{ -4 }$  \\
$ 21.6 - 21.7 $ & $ 9.48 \times 10^{ -4 }$ & $ [4.74  - 18.95 ]\times 10^{ -4 }$ & $0 -  2.37 \times 10^{ -3 }$  \\
$ 21.7 - 21.8 $ & $0$ & $0 -  7.53 \times 10^{ -4 }$ & $0 -  1.13 \times 10^{ -3 }$  \\
$ 21.8 - 21.9 $ & $ 2.99 \times 10^{ -4 }$ & $0 -  8.97 \times 10^{ -4 }$ & $0 -  1.20 \times 10^{ -3 }$  \\
$ 21.9 - 22.0 $ & $ 2.37 \times 10^{ -4 }$ & $0 -  7.12 \times 10^{ -4 }$ & $0 -  9.50 \times 10^{ -4 }$  \\
$ 22.0 - 22.1 $ & $ 3.77 \times 10^{ -4 }$ & $ [3.77  - 5.66 ]\times 10^{ -4 }$ & $ [1.89  - 7.55 ]\times 10^{ -4 }$  \\
$ 22.1 - 22.2 $ & $0$ & $0 -  3.00 \times 10^{ -4 }$ & $0 -  3.00 \times 10^{ -4 }$  \\
$ 22.2 - 22.3 $ & $0$ & $0 -  1.19 \times 10^{ -4 }$ & $0 -  2.38 \times 10^{ -4 }$  \\
$ 22.3 - 22.4 $ & $0$ & $0 -  1.89 \times 10^{ -4 }$ & $0 -  1.89 \times 10^{ -4 }$  \\
$ 22.4 - 22.5 $ & $0$ & $0 -  1.50 \times 10^{ -4 }$ & $0 -  1.50 \times 10^{ -4 }$  \\
$ 22.5 - 22.6 $ & $0$ & $0 -  5.97 \times 10^{ -5 }$ & $0 -  1.19 \times 10^{ -4 }$  \\
$ 22.6 - 22.7 $ & $0$ & $0 -  4.74 \times 10^{ -5 }$ & $0 -  9.48 \times 10^{ -5 }$  \\
$ 22.7 - 22.8 $ & $0$ & $0 -  3.76 \times 10^{ -5 }$ & $0 -  7.53 \times 10^{ -5 }$  \\
$ 22.8 - 22.9 $ & $0$ & $0 -  2.99 \times 10^{ -5 }$ & $0 -  2.99 \times 10^{ -5 }$  \\
\hline
\end{tabular}
 \caption{Column density distribution function for all DLAs with $4<z<5$. See also Figure \ref{fig:cddf_zz}.}
\label{tab:DR12/cddf_z45.txt}
 \end{table*}

\begin{table*}
 \centering
\begin{tabular}{ccccccccc}
\hline
$z$ & $\mathrm{d}N/\mathrm{d}X$ & $68$\% limits & $95$\% limits & $\Omega_\mathrm{DLA} (10^{-3}) $ & $68$\% limits & $95$\% limits \\
 \hline
$ 2.00 - 2.17 $ & $ 0.0540 $ & $ 0.0531 - 0.0550 $ & $ 0.0521 - 0.0559 $  & $ 0.783 $ & $ 0.738 - 0.834 $ & $ 0.698 - 0.887 $  \\
$ 2.17 - 2.33 $ & $ 0.0584 $ & $ 0.0575 - 0.0594 $ & $ 0.0565 - 0.0604 $  & $ 0.710 $ & $ 0.680 - 0.745 $ & $ 0.654 - 0.784 $  \\
$ 2.33 - 2.50 $ & $ 0.0584 $ & $ 0.0573 - 0.0595 $ & $ 0.0562 - 0.0606 $  & $ 0.744 $ & $ 0.706 - 0.789 $ & $ 0.676 - 0.836 $  \\
$ 2.50 - 2.67 $ & $ 0.0594 $ & $ 0.0582 - 0.0607 $ & $ 0.0569 - 0.0620 $  & $ 0.700 $ & $ 0.672 - 0.733 $ & $ 0.648 - 0.769 $  \\
$ 2.67 - 2.83 $ & $ 0.0712 $ & $ 0.0696 - 0.0728 $ & $ 0.0681 - 0.0744 $  & $ 0.784 $ & $ 0.755 - 0.816 $ & $ 0.729 - 0.849 $  \\
$ 2.83 - 3.00 $ & $ 0.0763 $ & $ 0.0745 - 0.0783 $ & $ 0.0727 - 0.0801 $  & $ 0.815 $ & $ 0.787 - 0.846 $ & $ 0.762 - 0.878 $  \\
$ 3.00 - 3.17 $ & $ 0.0813 $ & $ 0.0791 - 0.0836 $ & $ 0.0770 - 0.0858 $  & $ 0.946 $ & $ 0.910 - 0.985 $ & $ 0.878 - 1.024 $  \\
$ 3.17 - 3.33 $ & $ 0.0831 $ & $ 0.0803 - 0.0860 $ & $ 0.0776 - 0.0888 $  & $ 0.954 $ & $ 0.908 - 1.003 $ & $ 0.866 - 1.054 $  \\
$ 3.33 - 3.50 $ & $ 0.0903 $ & $ 0.0868 - 0.0943 $ & $ 0.0832 - 0.0979 $  & $ 1.094 $ & $ 1.001 - 1.211 $ & $ 0.925 - 1.343 $  \\
$ 3.50 - 3.67 $ & $ 0.0840 $ & $ 0.0797 - 0.0885 $ & $ 0.0757 - 0.0925 $  & $ 1.258 $ & $ 1.095 - 1.473 $ & $ 0.973 - 1.728 $  \\
$ 3.67 - 3.83 $ & $ 0.0742 $ & $ 0.0695 - 0.0798 $ & $ 0.0648 - 0.0849 $  & $ 1.104 $ & $ 0.846 - 1.391 $ & $ 0.694 - 1.679 $  \\
$ 3.83 - 4.00 $ & $ 0.0730 $ & $ 0.0662 - 0.0806 $ & $ 0.0601 - 0.0875 $  & $ 0.861 $ & $ 0.741 - 1.016 $ & $ 0.642 - 1.329 $  \\
$ 4.00 - 4.17 $ & $ 0.0796 $ & $ 0.0706 - 0.0899 $ & $ 0.0616 - 0.0988 $  & $ 1.108 $ & $ 0.906 - 1.329 $ & $ 0.730 - 1.563 $  \\
$ 4.17 - 4.33 $ & $ 0.1033 $ & $ 0.0894 - 0.1192 $ & $ 0.0775 - 0.1331 $  & $ 1.275 $ & $ 1.126 - 1.439 $ & $ 0.991 - 1.619 $  \\
$ 4.33 - 4.50 $ & $ 0.1203 $ & $ 0.1027 - 0.1408 $ & $ 0.0880 - 0.1584 $  & $ 2.335 $ & $ 1.609 - 3.608 $ & $ 1.250 - 5.201 $  \\
$ 4.50 - 4.67 $ & $ 0.1620 $ & $ 0.1358 - 0.1927 $ & $ 0.1139 - 0.2190 $  & $ 2.609 $ & $ 2.158 - 3.126 $ & $ 1.735 - 3.697 $  \\
$ 4.67 - 4.83 $ & $ 0.1881 $ & $ 0.1580 - 0.2333 $ & $ 0.1204 - 0.2709 $  & $ 1.920 $ & $ 1.510 - 2.370 $ & $ 1.150 - 2.839 $  \\
$ 4.83 - 5.00 $ & $ 0.0956 $ & $ 0.0638 - 0.1435 $ & $ 0.0319 - 0.1913 $  & $ 1.340 $ & $ 0.696 - 2.125 $ & $ 0.221 - 3.014 $  \\
\hline
  \end{tabular}
 \caption{Table of $\mathrm{d}N/\mathrm{d}X$ and $\Omega_\mathrm{DLA}$
    from our catalogue for $2 < z < 5$. See also Figure \ref{fig:omega_dla} and Figure \ref{fig:dndx}.}
\label{tab:DR12/dndx_all.txt}
 \end{table*}

\label{lastpage}
\bibliography{GPCDDF}

\end{document}